\documentclass[a4paper,12pt]{article}
\usepackage{amsmath,amssymb}
\usepackage{graphicx}
\usepackage{units}
\usepackage{thophys}
\DeclareMathOperator{\atan}{atan}

\title{%
  Production of Almost~Fermiophobic~Gauge~Bosons in the
  Minimal~Higgsless~Model at~the~LHC}

\author{%
  Thorsten Ohl%
   \thanks{\texttt{ohl@physik.uni-wuerzburg.de}}\\
  Christian Speckner%
   \thanks{\texttt{cnspeckn@physik.uni-wuerzburg.de}}\\
  \hfil\\
  Institut f\"ur Theoretische Physik und Astrophysik\\
  Universit\"at W\"urzburg\\
  Am Hubland, 97074 W\"urzburg, Germany}
\date{\today}

\begin{document}
\maketitle

\begin{abstract}
  We study the production of the heavy~$W'$ and $Z'$ bosons in the
  three site higgsless model at the LHC.  We focus on the $s$-channel
  production mode to estimate the prospects for measuring their
  suppressed couplings to standard model fermions.
\end{abstract}

\section{Introduction}

At the eve of data taking at the LHC, the electroweak standard
model~(SM) with a fundamental scalar Higgs doublet remains an
extremely successful effective description of all data collected in
particle physics experiments at colliders.  Nevertheless, the
microscopic dynamics of the electroweak symmetry breaking~(EWSB)
sector has not yet been tested directly.  Therefore, detailed studies
of the SM and realistic alternative scenarios for EWSB are an
essential part of the LHC experimental program.

In the past decade, additional dimensions of space-time at the
TeV-scale have become an important paradigm for electroweak~(EW) model
building.  Planck-scale extra dimensions have long been a solid
prediction of superstring theory, but they are outside of the
experimental range of collider experiments.  In contrast, TeV-scale
extra dimensions will be tested at the LHC.

Models with just one additional space dimension that have the geometry of
a five dimensional Anti-de Sitter space~($\text{AdS}_5$)~\cite{Randall:1999ee},
play a special role, because the conjectured AdS/CFT
correspondence~\cite{Maldacena:1997re} reveals such models as dual to conformal
field theories~(CFT) on the four dimensional boundary branes.  In
particular, a weakly interacting~$\text{AdS}_5$ model turns out to be
dual to a strongly interacting Technicolor~(TC) like model for EWSB.
If the conjectured AdS/CFT correspondence is exact, the extra
dimension can be viewed as a technically convenient description
of a strongly interacting dynamics.

Model building for EWSB with extra dimensions does not require them to
be continuous, instead they can be
deconstructed~\cite{ArkaniHamed:2001ca} as a discrete lattice with a
finite number of sites.  In this approach, the extra dimensions play a
metaphorical role as a organization principle for gauge theories with
large non-simple gauge groups and complicated matter representations,
similar to moose models.  It turns out that a minimal version of
warped higgsless models can be (de)constructed on just three lattice
sites in the extra dimension and is known as the Three Site Higgsless
Model~(3SHLM)~\cite{SekharChivukula:2006cg}.

In order to be compatible with the EW precision tests~(EWPT), any
additional heavy gauge bosons should couple weakly to the SM fermions.
The 3SHLM ensures this by ``ideal fermion
delocalization''~\cite{SekharChivukula:2006cg,SekharChivukula:2005xm}
and the predominant production mechanism at the LHC will be in vector
boson fusion~\cite{He:2007ge}.  However, it has been pointed out
recently~\cite{Abe:2008hb}, that the EWPT actually require a small,
but nonvanishing coupling of the heavy gauge bosons to SM fermions.
This allows their production in the $s$-channel at the LHC.  In fact,
a measurement of the relative strengths of the production mechanisms
for heavy vector bosons will be required to constrain higgsless models
of EWSB.

In this paper, we complement the existing phenomenological
studies~\cite{He:2007ge} of the 3SHLM by allowing for non ideal
delocalization and the production of~$W'$ and $Z'$~bosons in the
$s$-channel at LHC.  We perform parton level Monte Carlo studies to
identify the regions of parameter space where the coupling of the
$W'$~boson to SM fermions can be measured at the LHC.  We show how the
contributions from the nearly degenerate ~$W'$ and $Z'$~bosons can be
separated for this purpose.

This paper is organized as follows: in section~\ref{sec:3shl} we
review the features of the 3SHLM that are relevant for our
investigation.  In section~\ref{sec-cpl} we discuss the relevant
couplings, masses and widths that are used in our Monte Carlo
eventgenerator described in section~\ref{sec:whizard}.
In sections~\ref{sec-hzprod} and~\ref{sec-hwprod} we discuss our
results for the production of heavy~$Z'$ and $W'$~bosons, respectively.
Limitations arising from finite jet mass resolutions are described
in~\ref{sec-jetres}.  We conclude in section~\ref{sec:concl}.

\section{The Three Site Higgsless Model}
\label{sec:3shl}

\begin{figure}
\centerline{\includegraphics{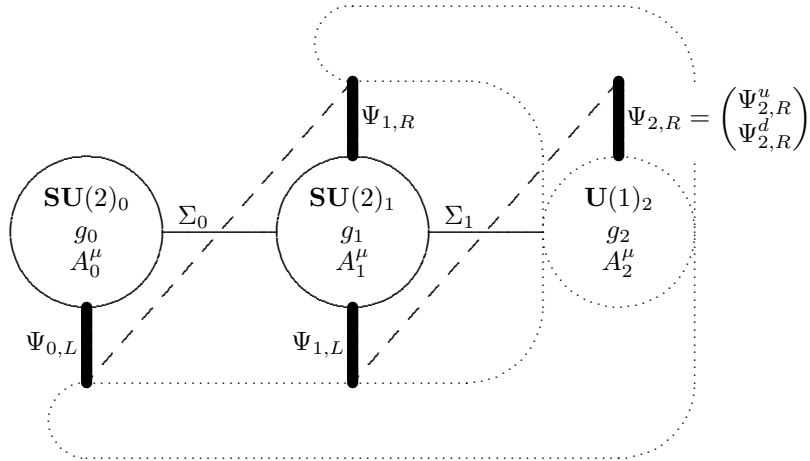}}
\caption{The field content and structure of the 3SHLM in moose notation. The dashed lines
connecting fermions represent Yukawa couplings, the dotted blob illustrates the nontrivial
$\mathbf{U}(1)_2$ charge carried by all fermions.}
\label{fig-moose}
\end{figure}

The Three Site Higgsless Model~\cite{SekharChivukula:2006cg} can be
viewed as a warped 5D model of EWSB,
dimensionally deconstructed~\cite{ArkaniHamed:2001ca}
to three lattice sites. The structure and field content of the model
is shown in moose notation in figure~\ref{fig-moose}.
The gauge group consists of two $\mathbf{SU}(2)$ group factors located at the
lattice sites~$0$ and~$1$ with gauge fields $A^\mu_{0/1}$ and gauge couplings $g_{0/1}$ and a
$\mathbf{U}(1)$ gauge group located at the third lattice site with the gauge field
$A^\mu_2$ and gauge coupling $g_2$.  Note that the continuous 5D
analogue of this is a bulk $\mathbf{SU}(2)$
broken to $\mathbf{U}(1)$ on one brane by boundary conditions. The lattice sites are
linked by $\mathbf{SU}(2)$ valued Wilson line fields $\Sigma_{0/1}$
that transform bi-unitarily under
gauge transformations as
\begin{equation*}
 \Sigma_0 \longrightarrow U_0\Sigma_0 U_1^\dagger \quad,\quad
 \Sigma_1 \longrightarrow U_1\Sigma_0 e^{-i\theta\frac{\sigma_3}{2}}\,.
\end{equation*}
If the potential for the Wilson line fields is arranged such that these acquire a vacuum
expectation value
\begin{equation*}
  \left<\Sigma_{0/1}\right> = \sqrt{2}v\,,
\end{equation*}
the symmetry group is broken spontaneously
to the electromagnetic~$\mathbf{U}(1)_\text{em}$. The kinetic terms for~$\Sigma_{0/1}$
contain covariant derivatives which produce mass terms for the gauge bosons;
after diagonalization we find a massless photon, two massive charged
gauge bosons~$W$ and~$W^\prime$ and two neutral massive gauge bosons~$Z$ and~$Z^\prime$. 

Choosing~$g_1\gg g_0,g_2$, the mass gap between the massive
gauge bosons becomes large and the lighter ones can be identified
with the SM~$W$  and $Z$~bosons. These are mostly localized at the
brane sites, while the heavy modes
are strongly localized at the bulk site. This symmetry breaking setup is similar to the
BESS model~\cite{Casalbuoni:1985kq}. After fixing the electric charge and
the~$W$ and $Z$~masses from the observed values,
the only remaining free parameter in the gauge sector is the $W^\prime$ mass~$m_{W'}$.

Fermions are incorporated into the model by putting left-handed $\mathbf{SU}(2)$
doublets $\Psi_{0/1,L}$ on the sites~$0$ and~$1$, a right-handed doublet $\Psi_{1,R}$ on
site~$1$ and singlets $\Psi^{u/d}_{2,R}$ on site~$2$ for every SM
fermion (cf.~figure~\ref{fig-moose}).  The $\mathbf{U}(1)_2$~charges
of the $\Psi^{u/d}_{2,R}$ fermions are taken from the SM hypercharge
assignments for the corresponding righthanded singlets, whereas the
$\mathbf{U}(1)_2$~charges of all other left- and righthanded fermions
are taken from the SM hypercharge assignments for the corresponding
\emph{left}handed doublets.

In addition to the kinetic terms,
Yukawa couplings are added to the fermion Lagrangian
\begin{multline}
\label{equ-lgr-fermion}
\mathcal{L}_\text{Yukawa} = \\
\lambda\sum_i
\left[\epsilon_L\overline{\Psi}_{0,L}^i\Sigma_0\Psi_{1,R}^i +
\sqrt{2}v\overline{\Psi}_{1,L}^i\Psi_{1,R}^i +
\overline{\Psi}_{1,L}^i\Sigma_1
  \begin{pmatrix}
     \epsilon_{u,R}^i & 0 \\
     0                & \epsilon_{d,R}^i
  \end{pmatrix}
\Psi_{2,R}^i\right]
\end{multline}
with the index~$i$ running over all SM fermions.
The parameter $\epsilon_L$ is chosen universally for all fermions and
such that the tree-level corrections to the EWPT
vanish.  This will be referred to as ``ideal fermion
delocalization''~\cite{SekharChivukula:2006cg,SekharChivukula:2005xm}.
The parameter $\lambda$
is also chosen universally for all fermions; only the~$\epsilon_{u/d,R}$ have a nontrivial
flavor structure and are used to implement the mass splitting of
quarks and leptons, as well as CKM flavor mixing. The
vacuum expectation value $v$ breaks the symmetry and the mass
eigenstates are the SM
fermions (localized mostly at the branes) and heavy partner fermions (localized
mostly in the bulk).

The only remaining free parameter in the fermion sector after fixing the SM fermion
masses and $\epsilon_L$ is the heavy fermion mass scale $m_\text{bulk}=\sqrt{2}\lambda v$.
Therefore the model is fixed uniquely by setting the SM parameters,
$m_{W'}$ and $m_\text{bulk}$ and 
by the requirement of ideal delocalization.
Loop corrections to the EWPT
and other phenomenological bounds limit the minimal values for these two parameters,
requiring $m_{W'}>\unit[380]{GeV}$ and
$m_\text{bulk}>\unit[2]{TeV}$~\cite{SekharChivukula:2006cg}.

The spectrum of the model consists of the SM gauge bosons and
fermions, the $W^\prime$ and $Z^\prime$ and a heavy partner fermion
for each SM fermion. The masses of the two new heavy gauge bosons are
quasi-degenerate ($|m_{W'}-m_{Z'}|=\mathcal{O}(\unit[1]{GeV})$), and the
masses of the partner fermions are of the order $m_\text{bulk}$ with
the~$t^\prime$ being slightly heavier than the rest.

\section{Couplings, widths and branching ratios}
\label{sec-cpl}

Ideal fermion delocalization implies that the couplings of the light SM fermions
to the $W^\prime$ vanish and that those to the $Z^\prime$ are small
($\mathcal{O}(10^{-2}))$. The both heavy gauge bosons also
couple to the SM $Z$ and $W$~bosons with couplings of order $\mathcal{O}(10^{-2})$.

\begin{figure}
\centerline{\includegraphics[angle=270,width=8cm]{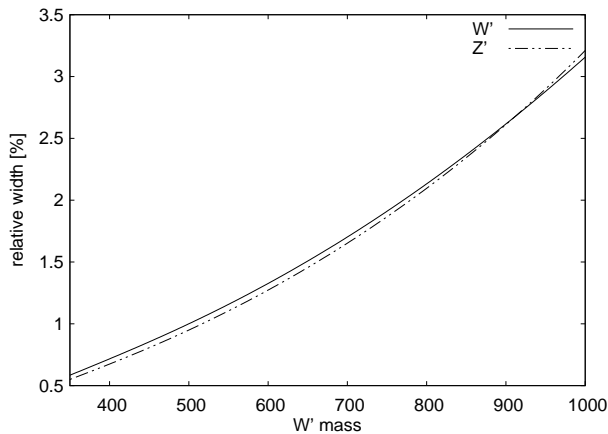}}
\caption{\label{plot-widths}%
The relative width~$\Gamma_V/m_V$ of the
$W^\prime$ and $Z^\prime$ bosons as a function of $m_{W'}$ with ideal
delocalization and $m_\text{bulk}=\unit[5]{TeV}$.}
\end{figure}
Therefore, the only decay channel for the $W^\prime$ in the ideally delocalized scenario is
the decay into a $W$ and a $Z$. The $Z^\prime$ can in principle also decay into SM
fermions; however, the decay of the longitudinal mode enhances the $WZ$ decay
channel by a factor of
\begin{equation*}
  \frac{m_{Z^\prime}^4}{16m_W^2m_Z^2}
\end{equation*}
over the decay into a fermion pair causing the latter decay to be highly
suppressed by a factor of the order of $\mathcal{O}(10^{-2})$ (cf.~\cite{He:2007ge}).
Looking at figure~\ref{plot-widths} we find that the resonances are rather narrow
($\Gamma_V/m_V\approx1-3\%$) improving the prospects for observing these particles at the LHC.

The new heavy fermions decay into their light partner and a gauge boson, the resulting
widths being of the order $\Gamma_f/m_f\approx0.1$, which, combined with their large mass
($>\unit[2]{TeV}$), will make the direct detection as a resonance at a collider rather
challenging.

For a massless SM fermion, the Yukawa coupling between the sites~$1$ and~$2$ vanishes.
From~$\mathcal{L}_{\text{Yukawa}}$ we find that the wave function is
completely fixed by the delocalization parameter $\epsilon_L$. Therefore, the influence of
$m_\text{bulk}$ on the wave functions of the light SM fermions and their couplings is very
small and the dependence of the cross section on $m_\text{bulk}$ is
almost negligible at LHC energies.

Although ideal delocalization guarantees compatibility with the
constraints from EWPT at tree level~\cite{SekharChivukula:2005xm},
a recent 1-loop analysis~\cite{Abe:2008hb} has shown that a deviation from
ideal delocalization is necessary to comply with the EWPT constraints at loop
level. According to the authors of~\cite{Abe:2008hb}, this deviation corresponds to an
on-shell coupling between the $W^\prime$ and the light fermions as large as $1-2\%$ of
the isospin gauge coupling $g_W\approx g_0$.

\begin{figure}
\centerline{\includegraphics{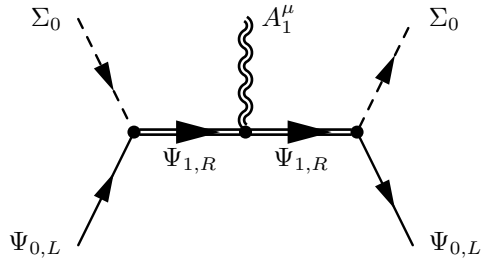}}
\caption{The tree-level diagram generating the operator~$O_1$~(cf.~(\ref{equ-op1})) after
integrating out the bulk fermions.}
\label{fig-op1}
\end{figure}
The coupling $g_{W'ff}$ to which the bounds derived in~\cite{Abe:2008hb} apply is defined
in the effective theory obtained by integrating out the bulk fermions and is renormalized
at the $W^\prime$ mass shell. There are two operators contributing to this coupling in the
one loop analysis in addition to the coupling of the left-handed
fermions to the component of
the $W^\prime$ sitting at site $0$. The first one
\begin{equation} 
\label{equ-op1}
  O_1 = \overline{\Psi}_{0,L}\Sigma_0\fmslash{A_1}\Sigma_0^\dagger\Psi_{0,L}
\end{equation}
encodes a coupling between the component of the $W^\prime$ sitting at site $1$ and the
left-handed SM fermion and is generated by integrating out the bulk fermion
from the diagram in figure~\ref{fig-op1} (see
also~\cite{SekharChivukula:2006cg}). The second operator
\begin{equation}
\label{equ-op2}
   O_2 = 
    \overline{\Psi}_{0,L}\left(D_\nu\left(\Sigma_0 F_1^{\mu\nu}\Sigma_0^\dagger\right)
      \right)\gamma_\mu\Psi_{0,L}
\end{equation}
arises from loop corrections. Although this operator also contains a contribution to the
coupling between the left-handed fermion and the gauge bosons at site $1$, it has a
nontrivial momentum structure. However, using a non-linear field redefinition in the
spirit of on-shell effective field theory~\cite{Georgi:1991ch}, the
corresponding part of~$O_2$ 
can be converted to the same form as~$O_1$ at the price of introducing additional
higher dimensional operators coupling at least two gauge bosons to two fermions whose
contributions are suppressed by another power of the gauge couplings. This allows the
operator~$O_2$ to be included into $g_{W^\prime ff}$ where it contributes to
the bounds derived by the authors of~\cite{Abe:2008hb}.

Therefore, the contributions of both these operators can be accounted for by adjusting the
delocalization parameter $\epsilon_L$ in the tree level
Lagrangian $\mathcal{L}_{\text{Yukawa}}$
to generate the coupling $g_{W^\prime ff}$. The model parameters then should be understood
to be renormalized at the $W^\prime$~mass.

\begin{figure}
\centerline{\includegraphics[angle=270,width=8cm]{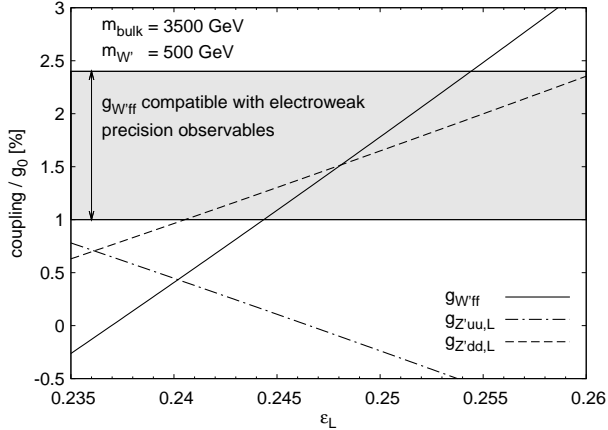}}
\caption{$g_{W'ff}$, $g_{Z'uu,L}$ and $g_{Z'dd,L}$ normalized to the site $0$ gauge
coupling as a function of the delocalization parameter $\epsilon_L$.
The gray rectangle marks the range for $g_{W^\prime ff}$ allowed by the EWPT as
derived by the authors of~\cite{Abe:2008hb}.}
\label{fig-gwff-500}
\end{figure}
In the case of light SM fermions and their partners, only the wave functions of the
left-handed fermions depend on the delocalization parameter $\epsilon_L$. Therefore, the
right-handed couplings between the new gauge bosons and the light SM fermions are not
affected by the departure from ideal delocalization. Denoting the wave functions
by~$\phi_{f,L,i}$ and~$\phi_{Z',i}$ and using the normalization of the fermion
wave functions, the left-handed coupling of a fermion
to the~$Z^\prime$ can be written as
\begin{equation}
  \sum_{i=0}^{1}\phi_{f,L,i}^2\left(\pm\frac{1}{2} g_i\phi_{Z',i}
      + Y g_2\phi_{Z',2}\right)
    = \pm\frac{1}{2}\sum_{i=0}^1 g_i\phi_{Z',i}\phi_{f,L,i}^2 + Yg_2\phi_{Z',2}\,,
\label{equ-cpl-zff}\end{equation}
with the sign depending on the isospin of the fermion and $Y$ denoting the hypercharge.
As tuning away from ideal delocalization shifts the light mode of the
fermion towards the heavy $Z^\prime$ sitting at site $1$, the isospin dependent part
in~(\ref{equ-cpl-zff}) grows, while the correction to~$g_{Z'uu}$ differs only in sign
from the correction to~$g_{Z'dd}$.

Figure~\ref{fig-gwff-500} shows the dependence of
$g_{W'ff}$, $g_{Z'dd}$ and $g_{Z'uu}$ on the delocalization parameter and clearly
demonstrates this behavior. Considering that we have both $u\bar u$
and $d\bar d$ initial states at
the LHC, that both couplings start with positive values of the same order of magnitude at
the point of ideal delocalization and that we also have right-handed couplings of the same
order of magnitude which don't depend on $\epsilon_L$ at all, we don't
expect a large impact from changing $\epsilon_L$ on $Z^\prime$~production in the $s$-channel.
On the other hand, the effect on the $s$-channel $W^\prime$~production
should be sizable, because $\epsilon_L$ interpolates between this channel
being forbidden and being about the same order of magnitude as
$Z^\prime$ production.
\begin{figure}
\centerline{\begin{tabular}{cc}
\includegraphics[angle=270,width=6.5cm]{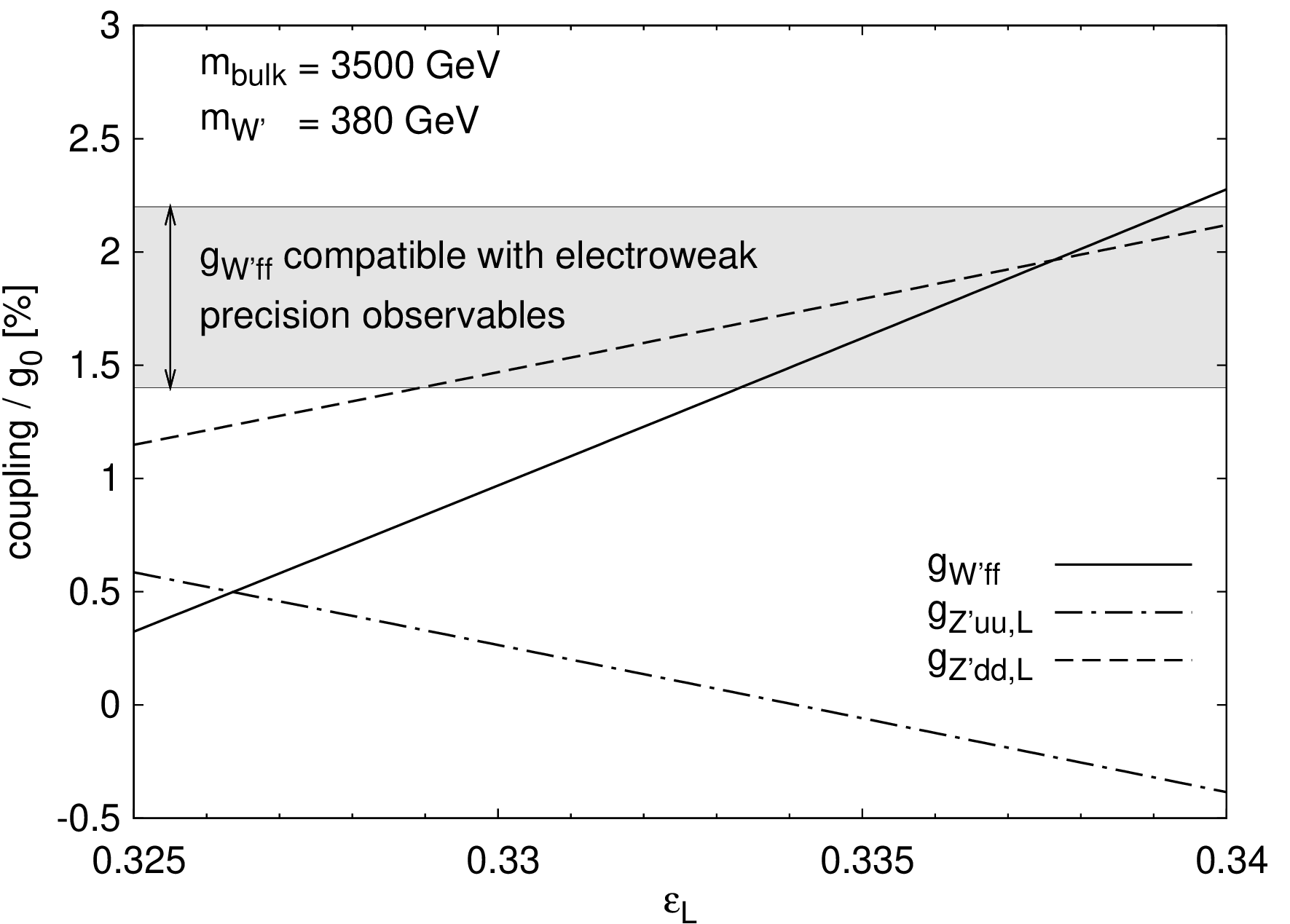} &
\includegraphics[angle=270,width=6.5cm]{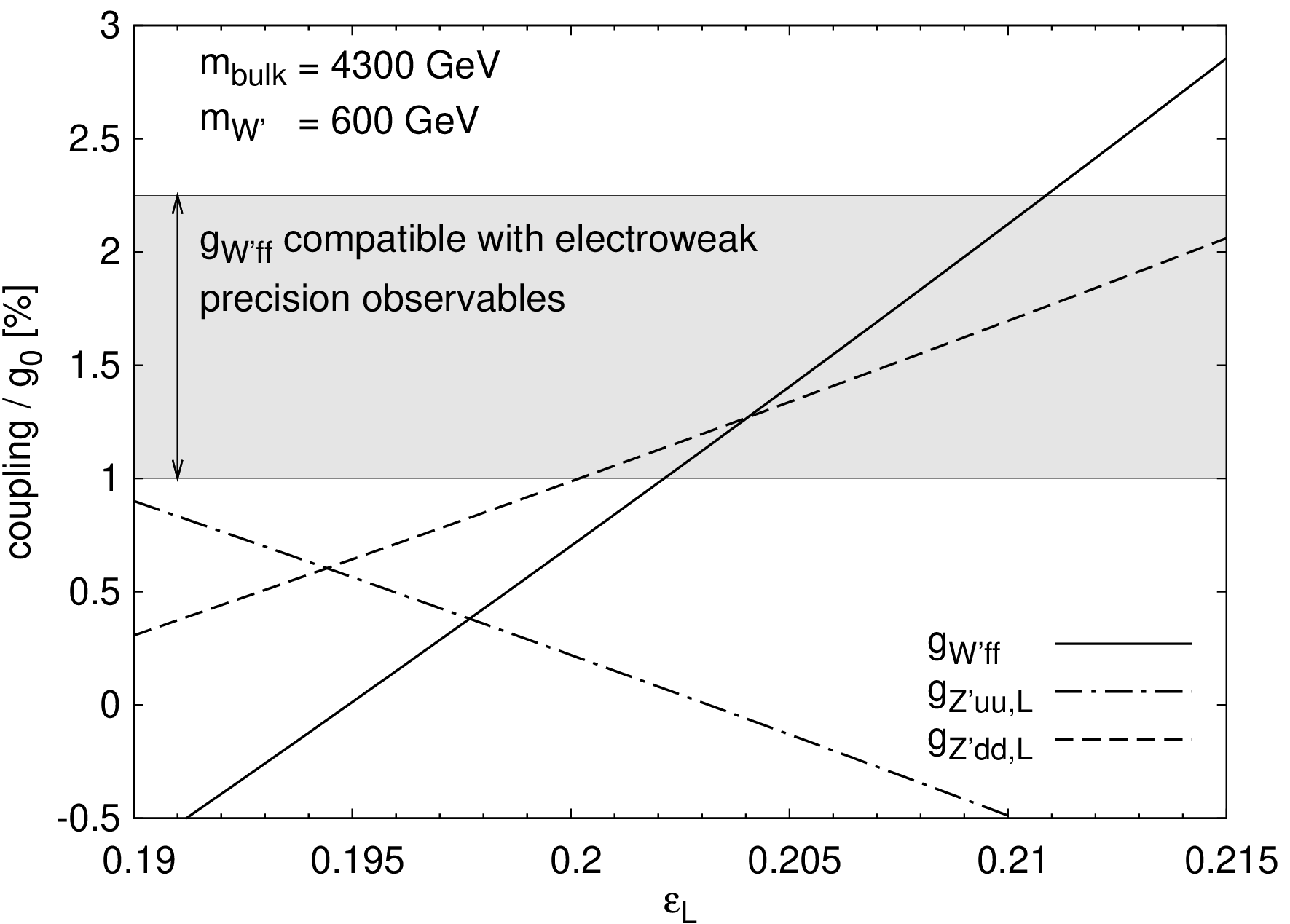}
\end{tabular}}
\caption{\label{fig-gwff-380-600}%
The same plots as figure~\ref{fig-gwff-500}, but for the other parts of parameter
space probed by our Monte Carlo simulations.}
\end{figure}

Figure~\ref{fig-gwff-380-600} shows the same plot as
figure~\ref{fig-gwff-500} for the other
regions of parameter space probed in our Monte Carlo simulations. In all three plots,
changing $m_\text{bulk}$ doesn't generate a visible change of the actual couplings, but
does move the gray band of acceptable $g_{W'ff}$ values.

\section{Implementation}
\label{sec:whizard}

We have coded a FORTRAN 90 module which diagonalizes the lagrangian of the model
and calculates all masses and couplings. Furthermore, the module calculates the tree
level widths of all new particles. Non ideal delocalization is implemented by tuning the
parameter $\epsilon_L$ away from the value required for vanishing $g_{W'ff}$.
For the automatized generation of tree level matrix
elements, we encoded the model in unitarity gauge into the optimizing
matrix element generator O'Mega~\cite{Moretti:2001zz,Kilian:2007gr}
which is part of the Monte Carlo eventgenerator generator WHIZARD~\cite{Kilian:2007gr}.
The results presented below are based on Monte
Carlo simulations using WHIZARD~\cite{Kilian:2007gr}.

We checked the couplings calculated by our FORTRAN code against all the couplings for
which analytic expressions are given in~\cite{SekharChivukula:2006cg}. To check the
validity of our implementation of the model, we compared the cross sections for a number of
$2\rightarrow 2$ processes to the SM, taking $m_{W^\prime}$,
$m_\text{bulk}$ and $m_\text{Higgs}$ to be huge. The widths calculated by the FORTRAN
module using analytic formulae were checked against numeric results obtained from
amplitudes generated by O'Mega.

We also checked gauge invariance by numerically checking the Ward Identities in the model 
obtained by taking the limit
\begin{equation*}
  \sqrt{2}v=\left<\Sigma_{0/1}\right>\rightarrow 0\,,
\end{equation*}
where the exact $\mathbf{SU}(2)_0\times\mathbf{SU}(2)_1\times\mathbf{U}(1)_2$
gauge symmetry is restored.

In addition, we compared several $2\rightarrow 2$ cross sections
to the CalcHep~\cite{Pukhov:2004ca} implementation of the model used by the authors
of~\cite{He:2007ge}. After plugging in the correct $W^\prime$ and $Z^\prime$ widths,
the results turn out to be in perfect agreement.

\section{$Z^\prime$ production in the $s$-channel}
\label{sec-hzprod}

In the ideally delocalized scenario, only the $Z^\prime$ has nonvanishing tree level
couplings to the SM fermions, while the $W^\prime$ is perfectly
fermiophobic. As explained above, the $Z^\prime$ decays with a branching ratio of over
$95\%$ into a $W^+W^-$ pair, rendering the resulting four fermion final state highly
favored over the two lepton one. This is in sharp contrast to many new heavy neutral
gauge bosons predicted by other extensions of the SM (Little Higgs, GUTs
etc.) which usually have larger fermion couplings but small or vanishing couplings to the
SM gauge bosons, because they typically originate from different gauge group factors and have little
or no mixing with the SM gauge bosons~\cite{Rizzo:2006nw,Langacker:2008yv}.

The most interesting final states for $Z^\prime$ production are thus $jjjj$, $l\nu jj$ and
$l\nu l\nu$. The four jet final state however is highly contaminated from
backgrounds containing gluon jets, and the two neutrino final state suffers from
the momentum information missing for the two neutrinos, leaving $l\nu jj$ as
the most promising candidate assuming one can cope with the missing neutrino momentum.
\begin{figure}
\centerline{\includegraphics{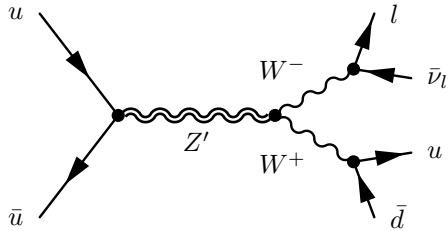}}
\caption{Representative of the class of diagrams contributing to the $Z^\prime$
production signal in $pp\rightarrow l\nu jj$.}
\label{fig-diag-hzprod}
\end{figure}
Figure~\ref{fig-diag-hzprod} shows a representative of the class of diagrams contributing to
the signal in this process.
In addition to the signal, there are also reducible backgrounds from events with
neutral jet pairs and an irreducible background from diagrams not of the type
figure~\ref{fig-diag-hzprod} contributing to the same final state.  In
this and the next section, we assume that a veto on forward tagging
jets is effective in suppressing the background from vector boson fusion.

\begin{figure}
\centerline{\includegraphics[angle=270,width=8cm]{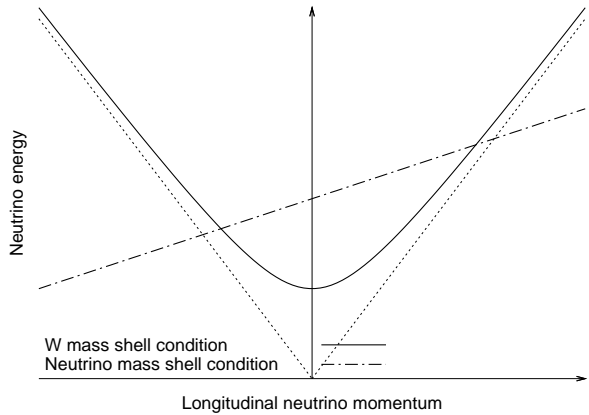}}
\caption{The two curves generated by the mass shell conditions for $W$ and neutrino in the
case of a $W$ decaying to $l\nu_l$.}
\label{fig-nurec}
\end{figure}

For the construction of an observable that can deal with the missing longitudinal neutrino
momentum, consider the decay of an on-shell $W$ into a lepton with momentum $p_l=q$ and a
neutrino with momentum $p_\nu=p$. The mass shell conditions of neutrino and $W$ boson then give
two equations involving the neutrino energy $p_0$ and longitudinal
momentum~$p_L$
\begin{subequations}
\begin{align}
\label{equ-nurec1}
  p_0^2 - p_L^2 - \left|\vec p_\perp\right|^2 &= 0 \\
\label{equ-nurec2}
  p_0 q_0 - p_L q_L - \vec p_\perp \vec q_\perp &= \frac{m_W^2}{2}
\end{align}
\end{subequations}
(assuming the lepton to be massless), with~$\vec p_\perp$ and~$\vec
q_\perp$ the projections of the momenta onto the transverse plane.
(\ref{equ-nurec1}) describes a hyperbola in the $p_L-p_0$
plane and~(\ref{equ-nurec2}) describes a straight line with the modulus of the slope
smaller than $1$. These curves are parametrized by $\vec p_\perp$, $q$ and $m_W$ and one
of their (two in general) intersections gives the neutrino energy and longitudinal
momentum as a function of these quantities. This geometrical situation is depicted in
figure~\ref{fig-nurec}.

This construction allows us to reconstruct the full neutrino momentum from the lepton
momentum and the missing $p_T$ for the events coming from the decay of a quasi-on-shell
$W$. However, owing to the modulus of the slope of the straight line being smaller than
one, we always have two solutions, none of which is preferred on kinematical
grounds. We have elected to deal with this by counting \emph{both} solutions in the histograms,
effectively doubling the amount of background events while preserving the size of the signal.
The two points of intersection can be obtained analytically by
\begin{equation}
p_0 = \frac{q_0^2\left(m_W^2 + 2\vec{p}_\perp\vec{q}_\perp\right) \pm q_L A}
	{2q_0\left(q_0^2  - q_L^2\right)} \quad,\quad
p_L = \frac{q_L\left(m_W^2 + 2\vec{p}_\perp\vec{q}_\perp\right) \pm A}
	{2\left(q_0^2 - q_L^2\right)}\,,
\label{equ-nurec3}
\end{equation}
with the abbreviation
\[
A = q_0\sqrt{\left(m_W^2 + 2\vec{p}_\perp\vec{q}_\perp\right)^2 +
	4\vec{p}_\perp^2\left(q_L^2 - q_0^2\right)} \,.
\]

To investigate the possibility of discovering the $Z^\prime$ in
$pp\rightarrow jjl+p_{T,\text{miss}}$ at the LHC we have performed full parton-level Monte
Carlo simulations for an integrated luminosity of $\int\mathcal{L}=\unit[100]{fb^{-1}}$,
the lepton being either an electron or a
muon and each jet being either a quark (excluding the top) or a gluon. To suppress the
backgrounds, we have applied $p_T$-cuts to all visible particles and to
$p_{T,\text{miss}}$
\[ p_T \ge \unit[50]{GeV}\,. \]
In addition, we have required the polar and intermediary angles of all visible particles
to lie within
\[ -0.95 \le \cos\theta \le 0.95 \]
and also applied a small-$x$ cut to the ingoing partons
\[ x \ge 1.4\cdot 10^{-3} \]
to avoid infrared singularities in the amplitude. For identifying the intermediary $W$ we
applied a cut to the invariant mass of the two jets\footnote{See
  section \ref{sec-jetres} for a discussion of the effects of finite jet 
  resolution on the identification of the $W$.}
\[ \unit[75]{GeV} \le m_{jj} \le \unit[85]{GeV}\,. \]
We used~(\ref{equ-nurec3}) to reconstruct the neutrino momentum, counting both solutions
into the histograms and discarding those with negative neutrino energy.

\begin{figure}
\centerline{\begin{tabular}{cc}
\includegraphics[angle=270,width=6.5cm]{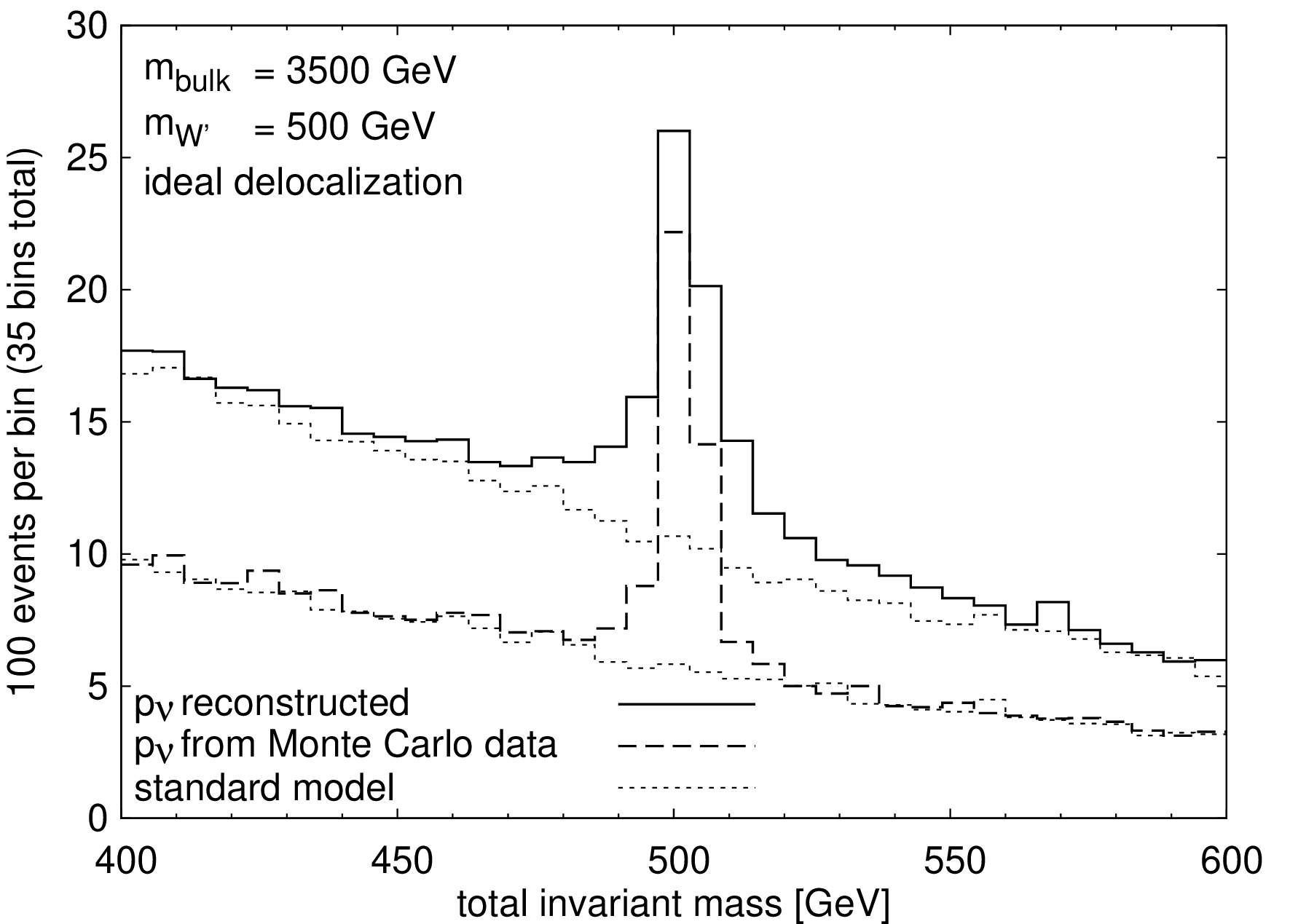} &
\includegraphics[angle=270,width=6.5cm]{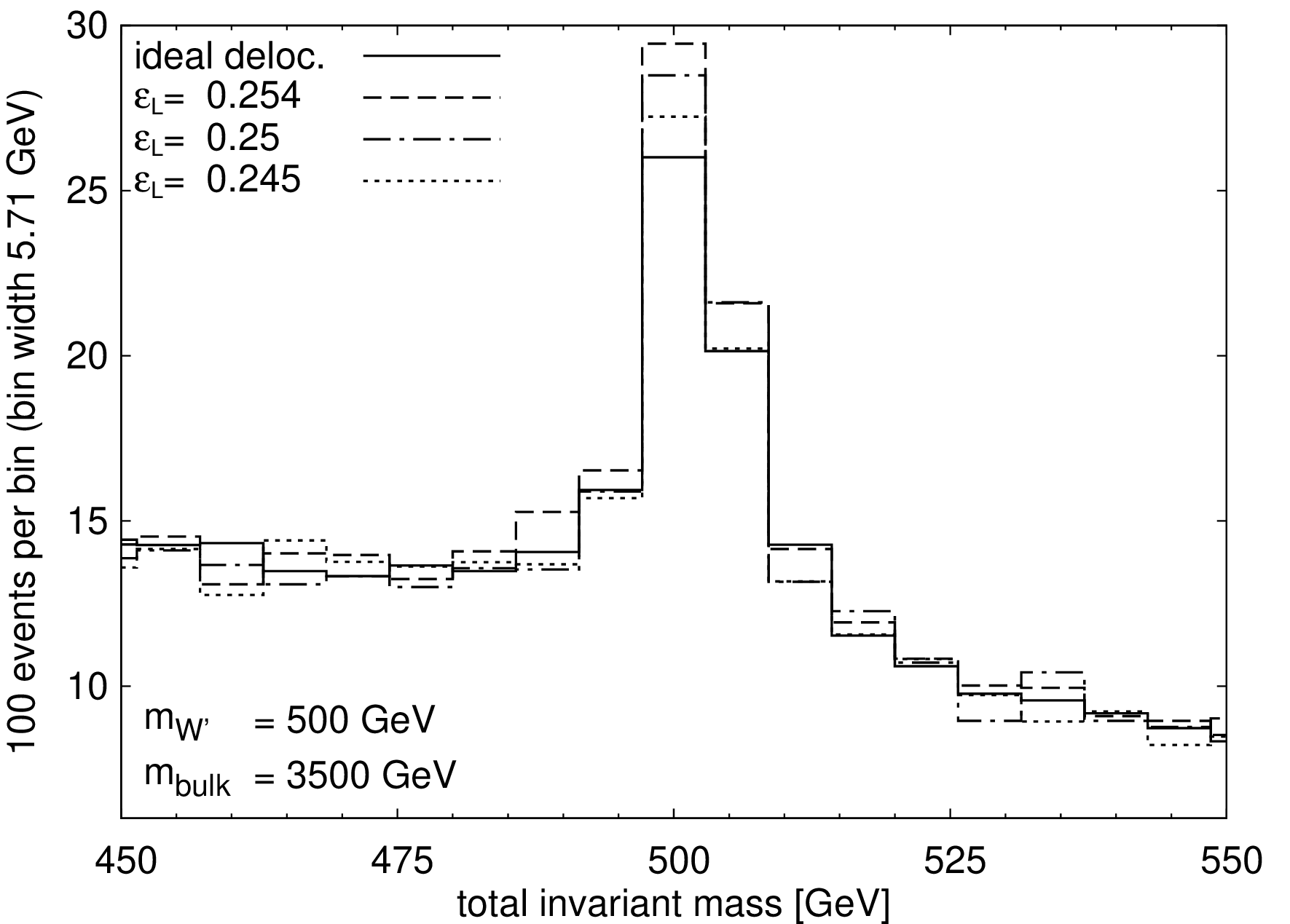}
\end{tabular}}
\caption{\emph{Left:} Invariant mass distribution in $pp\rightarrow l\nu_ljj$ obtained
from the reconstructed neutrino momenta vs. the distribution obtained from $p_\nu$
taken from Monte Carlo data. \emph{Right:} The effect of tuning $\epsilon_L$ away from
ideal delocalization (cf.~figure~\ref{fig-gwff-500}).}
\label{hist-hzprod-nurec}
\end{figure}
The plot on the left of figure~\ref{hist-hzprod-nurec} compares  the
invariant mass distribution obtained
from the reconstructed neutrino momenta to that obtained from
the unobservable neutrino momenta taken from Monte Carlo
data for $m_{W^\prime}=\unit[500]{GeV}$ and $m_\text{bulk}=\unit[3.5]{TeV}$.
In both cases the peak from the $Z^\prime$ is clearly visible. As expected, counting
both solutions obtained from the reconstruction doubles the amount of background events
while the number of events contained in the peak stays roughly the same. However, the peak is
broadened by the reconstruction, which can been seen when comparing to a
SM simulation (dotted line). The broadening at the center of the peak is mainly
caused by the mismatch between reconstructed and true neutrino momentum of the signal events
due to the $W$ not being exactly on-shell; the sidebands of the peak are caused by the
second solutions for $p_\nu$ of events at the center of the peak.

The plot on the right of figure~\ref{hist-hzprod-nurec} shows the
effect of changing the delocalization parameter
$\epsilon_L$ in the range allowed by the EWPT at one loop
(cf.~section \ref{sec-cpl}),
again for $m_{W^\prime}=\unit[500]{GeV}$ and $m_\text{bulk}=\unit[3.5]{TeV}$. As argued
before, the impact on the invariant mass distribution is not strong, the peak
staying clearly visible over the whole range of allowed values of $\epsilon_L$.

\begin{figure}
\centerline{\includegraphics[angle=270,width=8cm]{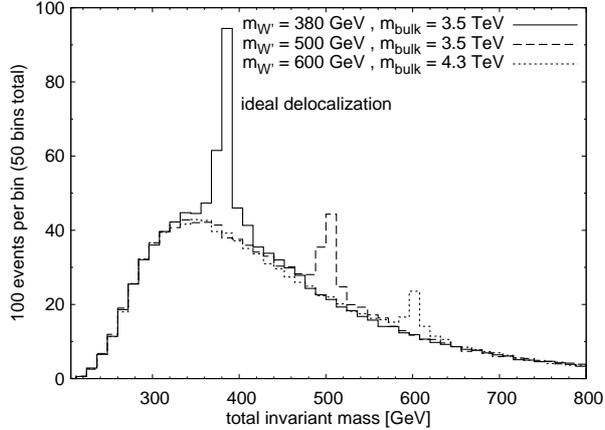}}
\caption{Invariant mass distribution in $pp\rightarrow l\nu_ljj$ for different values of
$m_{W^\prime}$ and $m_\text{bulk}$.}
\label{hist-hzprod-fullrange}
\end{figure}
Figure~\ref{hist-hzprod-fullrange} shows the invariant mass distributions obtained for
\[ m_{W^\prime}\in\left\{\unit[380]{GeV},\unit[500]{GeV},\unit[600]{GeV}\right\}\,,\]
which covers the whole range of values allowed by the EWPT
at one loop\footnote
{We also changed $m_\text{bulk}$ as shown in figure~\ref{hist-hzprod-fullrange} to
comply with the EWPT; however, as explained in section \ref{sec-cpl}, this has no
noticeable effect on the cross section.}~\cite{SekharChivukula:2006cg,Abe:2008hb}. As
the masses of the $Z^\prime$ and $W^\prime$ are quasi-degenerate,
the $Z^\prime$ peak moves with
changing $m_{W^\prime}$. The histogram shows that the peak stays clearly observable,
although it decreases in size as $m_{W^\prime}$ becomes larger owing
to the smaller parton
distribution functions for the sea quarks at larger values of~$x$.

To get a quantitative handle on the significance of the signal and to estimate the minimal
luminosity necessary for discovering the $Z^\prime$, we define the raw signal $N$ to be the
number of events in the $\pm\unit[20]{GeV}$ region around the peak. To estimate the
background we have generated SM events for an integrated luminosity of
$\int\mathcal{L}=\unit[400]{fb^{-1}}$, analyzed this data the in the same way
as the Monte Carlo data for the three site model and then downscaled the resulting
distributions by a factor of $4$ to reduce the error coming
from fluctuations in the background. We denote the number of background events in the
$\pm\unit[20]{GeV}$ region around the peak obtained this way by $N_b$.

We define the signal $N_s$ as
\begin{equation}
   N_s = N - N_b\,.
\end{equation}
The number of background events in the original Monte Carlo data $N^\prime_b$
is roughly doubled by our momentum reconstruction
\begin{equation*}
  N_b = 2N^\prime_b
\end{equation*}
and the standard deviation of $\sigma_{N_b}$ of $N_b$ must scale
accordingly, resulting in
\begin{equation*}
 \sigma_{N_b} = 2\sigma_{N^\prime_b} = 2\sqrt{N^\prime_b} = \sqrt{2N_b}\,.
\end{equation*}
We then define the significance in the usual way:
\begin{equation}
s = \frac{N_s}{\sigma_{N_b}} = \frac{N - N_b}{\sqrt{2N_b}}\,.
\label{equ-sgn-rec}\end{equation}
\begin{figure}
\centerline{\includegraphics[angle=270,width=8cm]{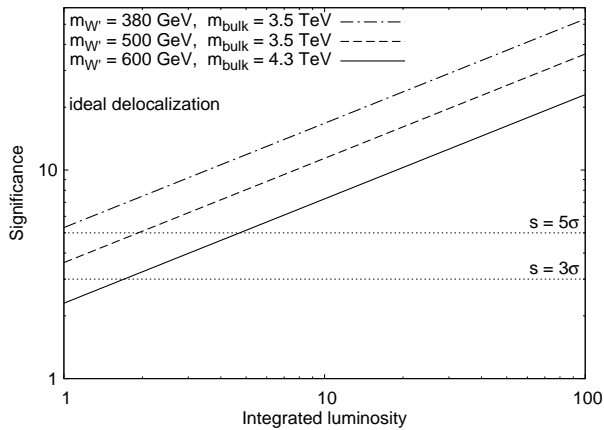}}
\caption{The significance as defined in the text as a function of the integrated
luminosity. The dotted lines mark the $3\sigma$ resp. $5\sigma$ discovery thresholds.}
\label{fig-sig-hz}
\end{figure}
The significance of the signal in the ideally delocalized scenario thus calculated is
shown in figure~\ref{fig-sig-hz} together with the $5\sigma$ and $3\sigma$ discovery
thresholds. The $5\sigma$ thresholds are approx.~$\unit[1]{fb^{-1}}$,
$\unit[2]{fb^{-1}}$, $\unit[5]{fb^{-1}}$ for
$m_{W^\prime}=\unit[380]{GeV}$, $\unit[500]{GeV}$, $\unit[600]{GeV}$, respectively.
Considering the fact that tuning
$\epsilon_L$ into the region allowed by the EWPT does not
significantly change the signal,
the three-site $Z^\prime$ may be discovered as early as in the first
$\unit[1-2]{fb^{-1}}$ and even in the worst case can be expected to
manifest itself in the first
$\unit[10-20]{fb^{-1}}$ of data.

\section{$W^\prime$ production in the $s$-channel without ideal delocalization}
\label{sec-hwprod}

\begin{figure}
\centerline{\begin{tabular}{cc}
\includegraphics{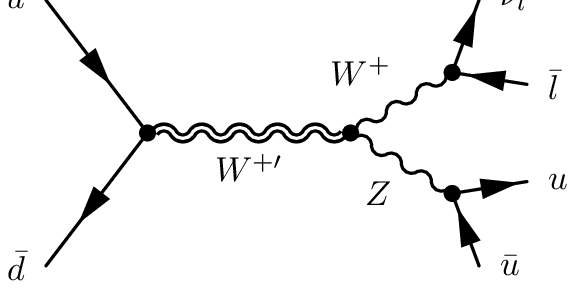} &
\includegraphics{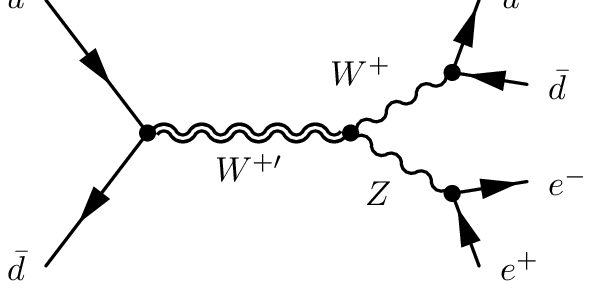}
\end{tabular}}
\caption{\emph{Left: }Representative of the class of diagrams contributing to the
$W^\prime$ production signal in $pp\rightarrow l\nu_ljj$. \emph{Right: }One of the signal
diagrams in the $lljj$ decay channel of the $W^\prime$.}
\label{fig-diag-hwprod}
\end{figure}
As discussed in section \ref{sec-cpl}, the deviation from ideal delocalization required by
the EWPT at one loop leads to non-vanishing couplings of the
$W^\prime$ to the SM fermions of the same order of magnitude as the $Z^\prime ff$
couplings. This allows for the possibility of producing the $W^\prime$ in the $s$-channel
at the LHC.

There are two possible decay channels for the $W^\prime$ that are promising candidates for
discovering this resonance. The first possibility is the decay $W^\prime\rightarrow
WZ\rightarrow l\nu_ljj$ (cf.~the left plot in figure~\ref{fig-diag-hwprod}),
which is the final state already discussed in the last section
and which can be treated the same way (replacing the cut on the $W$ mass with a cut on the
$Z$ mass). The second possibility is the decay of the $ZW$ pair into two leptons and two
jets (cf.~the right plot in figure~\ref{fig-diag-hwprod}). The absence
of missing $p_T$ is a clear advantage
of this decay mode allowing for background suppression by cutting on the invariant mass of
the lepton pair; unfortunately, the branching ratio is smaller than that for the $l\nu_ljj$
mode.

To probe the $l\nu_ljj$ final state we have used the same Monte Carlo data and cuts as in
section \ref{sec-hzprod} replacing the cut on the invariant mass\footnote{See
  section~\ref{sec-jetres} for a discussion of the effects of finite jet 
  resolution on the separation of $W$ and $Z$.}
of the jet pair with
\[ \unit[86]{GeV} \le m_{jj} \le \unit[96]{GeV}\,. \]
For probing the $lljj$ final state we again performed Monte Carlo simulations
for an integrated luminosity of $\int\mathcal{L}=\unit[100]{fb^{-1}}$. We applied the
same $p_T$, $x$ and angular cuts as in the last section together with the
identification cuts
\begin{equation}
\unit[75]{GeV} \le m_{jj} \le \unit[85]{GeV} \quad,\quad
\unit[86]{GeV} \le m_{ll} \le \unit[96]{GeV}
\label{equ-cut-wzident}\end{equation}
on the invariant mass of the jet pair and on that of the dilepton system.
\begin{figure}
\centerline{\begin{tabular}{cc}
\includegraphics[angle=270,width=6.5cm]{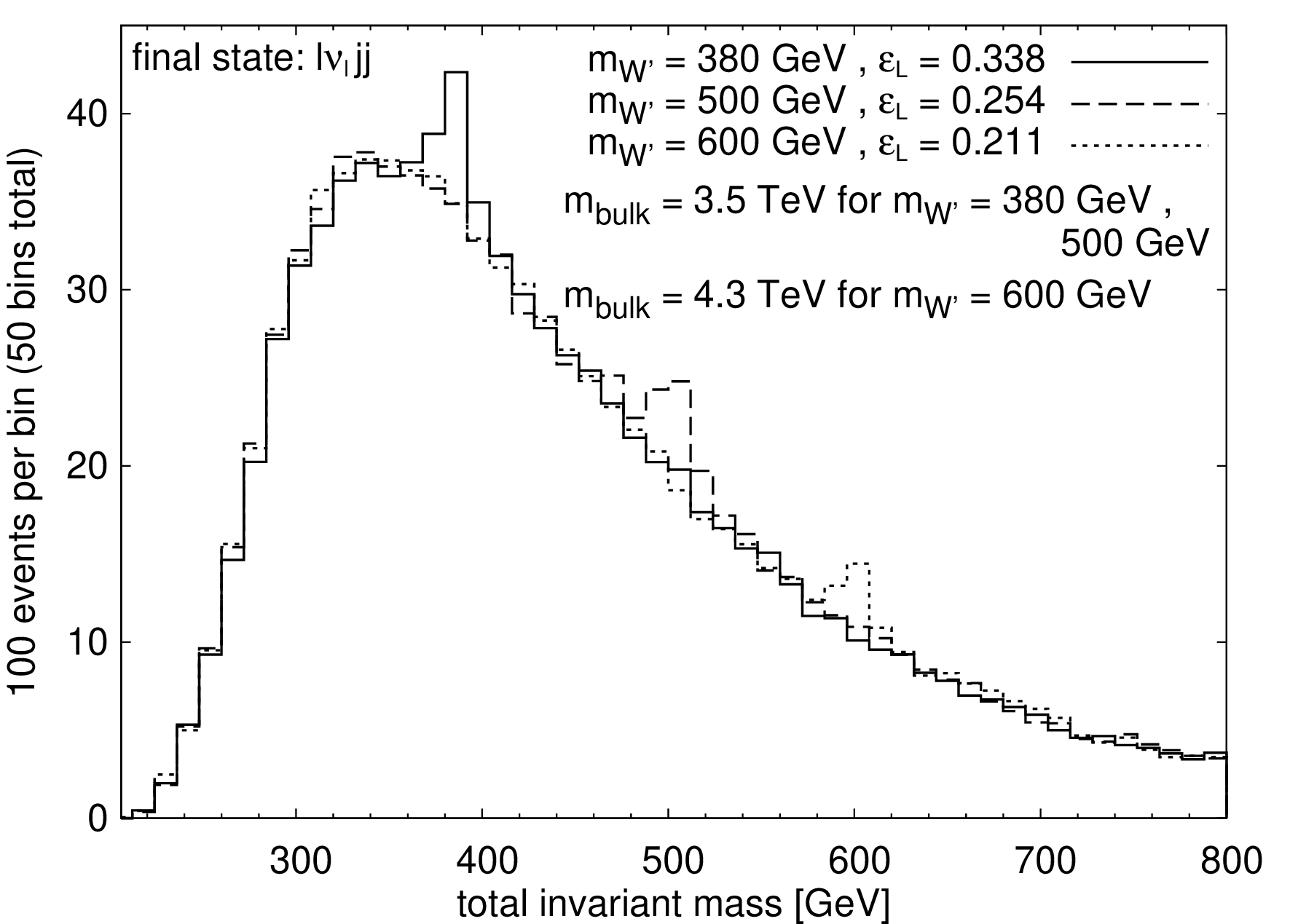} &
\includegraphics[angle=270,width=6.5cm]{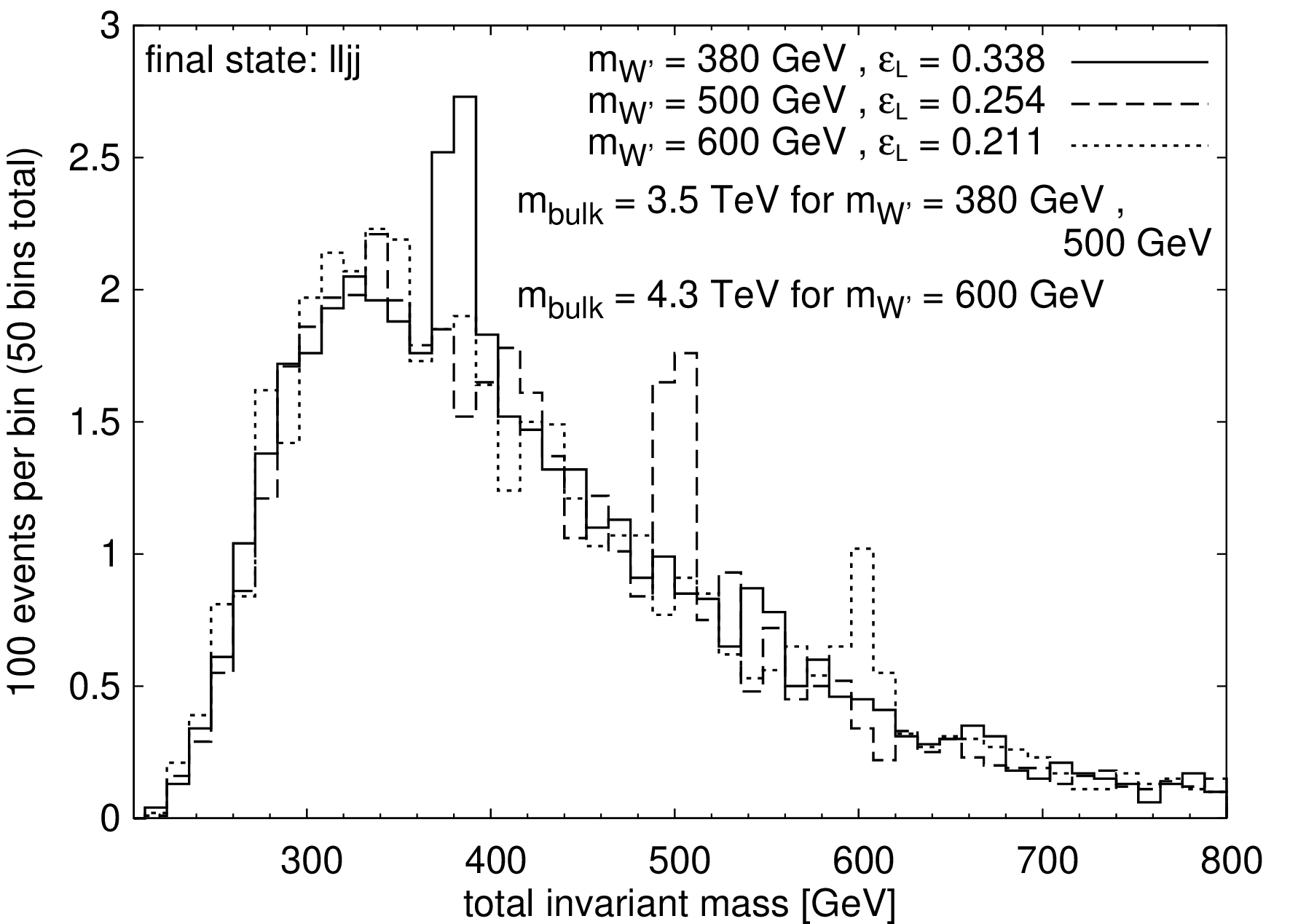}
\end{tabular}}
\caption{\emph{Left: }Invariant mass distribution for $W^\prime$ production in
$pp\rightarrow l\nu_ljj$ for different $W^\prime$ masses and large $g_{W'ff}$.
\emph{Right: }The same distribution for the $lljj$ final state.}
\label{hist-hw-fullrange}
\end{figure}
Figure~\ref{hist-hw-fullrange} shows the invariant mass distributions obtained for both final
states for $m_{W^\prime}=\unit[380]{GeV}$, $\unit[500]{GeV}$, $\unit[600]{GeV}$ and
$\epsilon_L$ chosen from the allowed range such as to give large
values\footnote{Even larger values of $g_{W'ff}$ are allowed by
  increasing $m_\text{bulk}$, but we are more interested in the lowest
  possible value for which the $W^\prime$ might still be detected in
  this channel at the LHC.}
of~$g_{W'ff}$ (cf.~figures~\ref{fig-gwff-500} and~\ref{fig-gwff-380-600}). For both final
states, the resonance peaks can be clearly seen for all three values of $m_{W^\prime}$.
The total number of events for $lljj$ is much smaller compared to $l\nu_l jj$
owing to the smaller branching
ratio, but the cuts on both $m_Z$ and $m_W$ and the absence of the double counting
introduced by the neutrino reconstruction significantly improve the
signal to background ratio.

\begin{figure}
\centerline{\begin{tabular}{cc}
\includegraphics[angle=270,width=6.5cm]{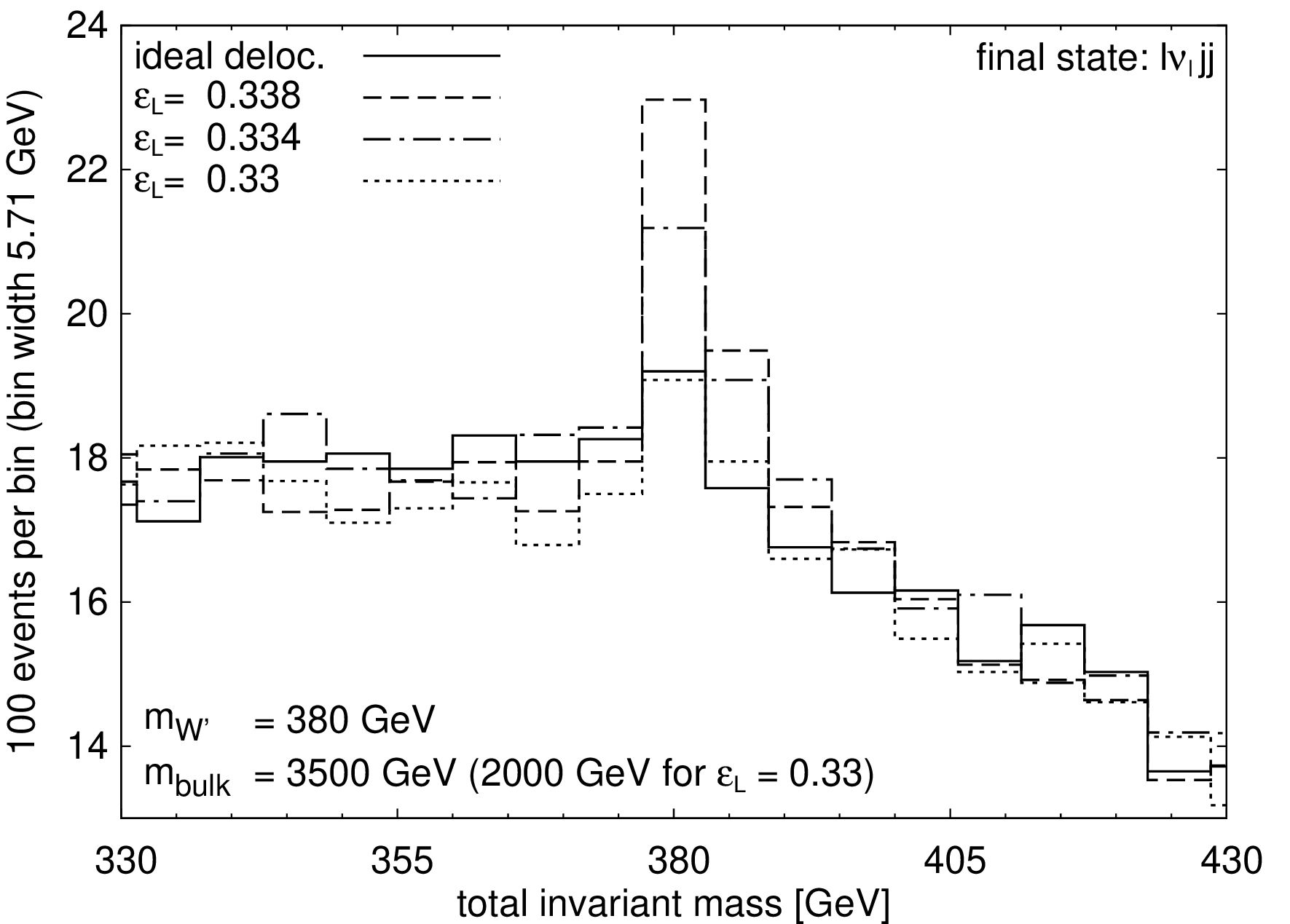} &
\includegraphics[angle=270,width=6.5cm]{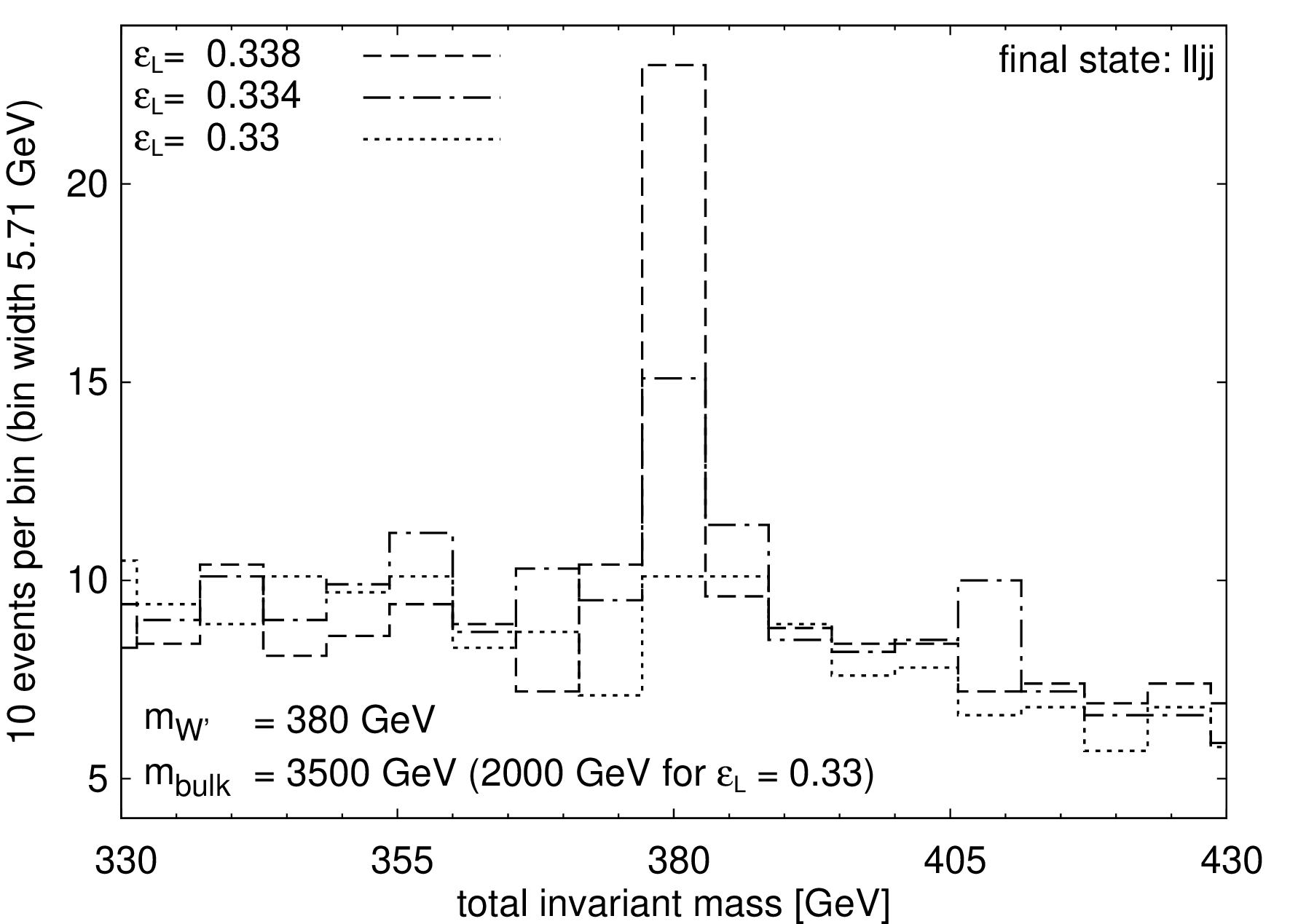} \\
\includegraphics[angle=270,width=6.5cm]{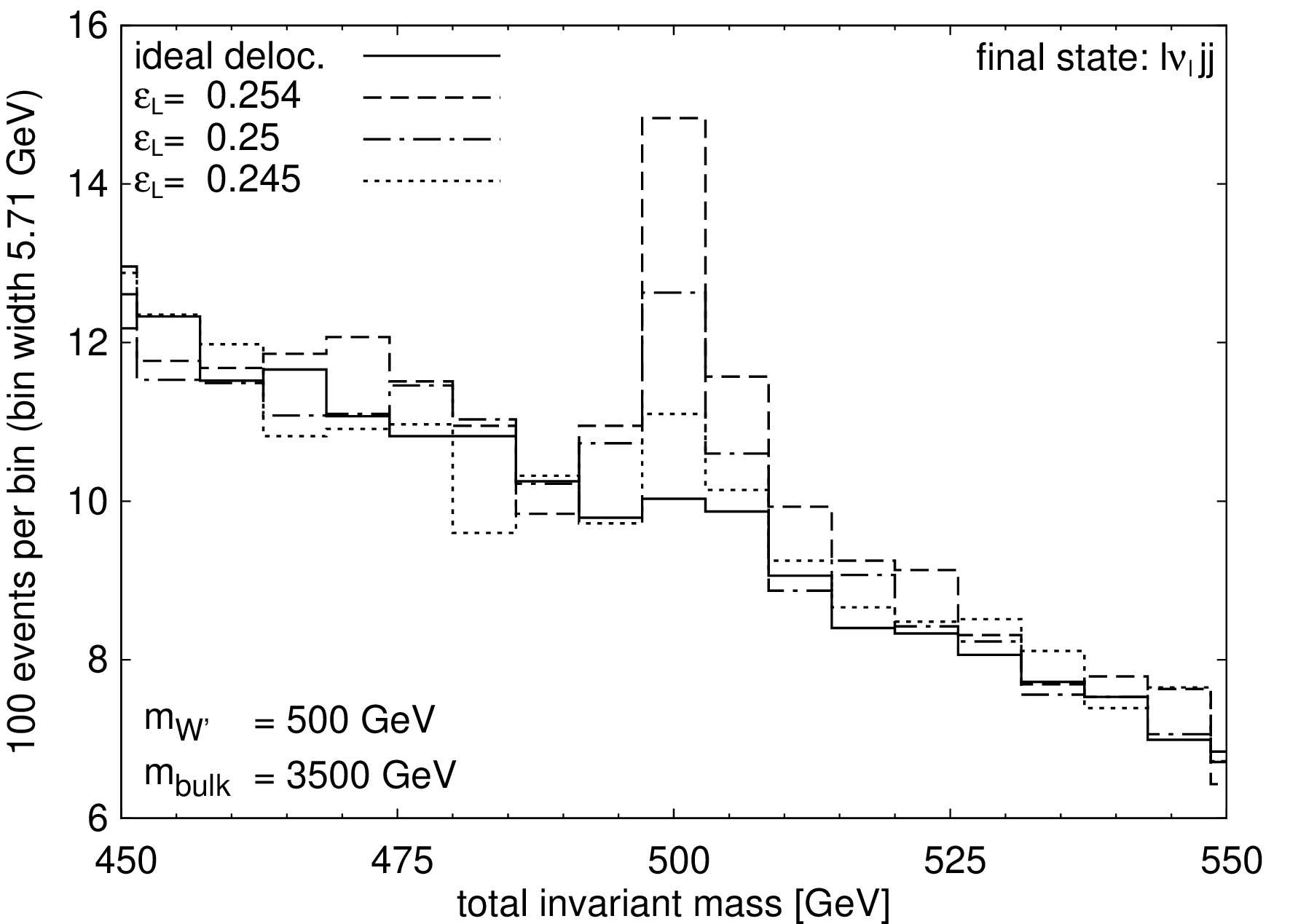} &
\includegraphics[angle=270,width=6.5cm]{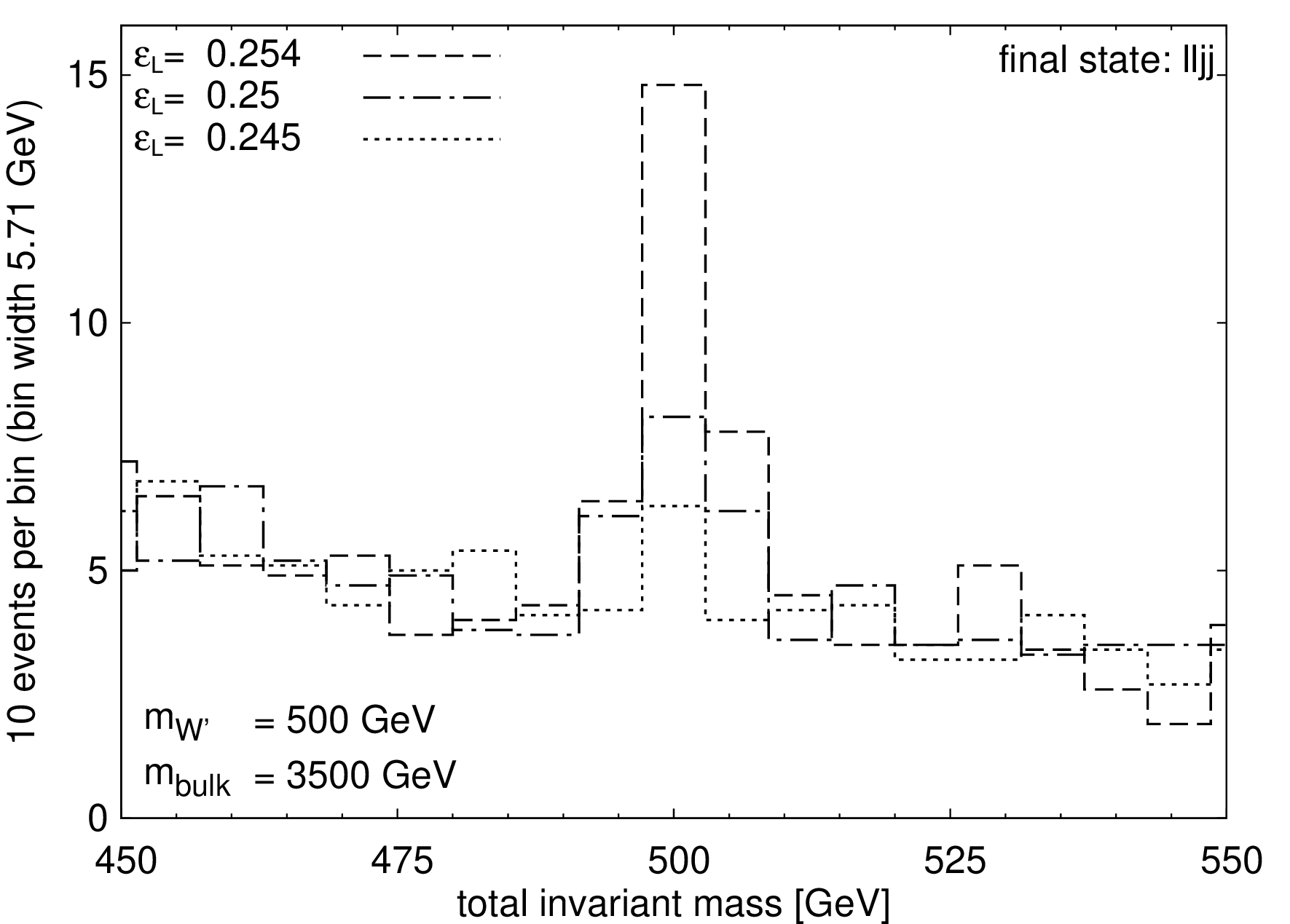} \\
\includegraphics[angle=270,width=6.5cm]{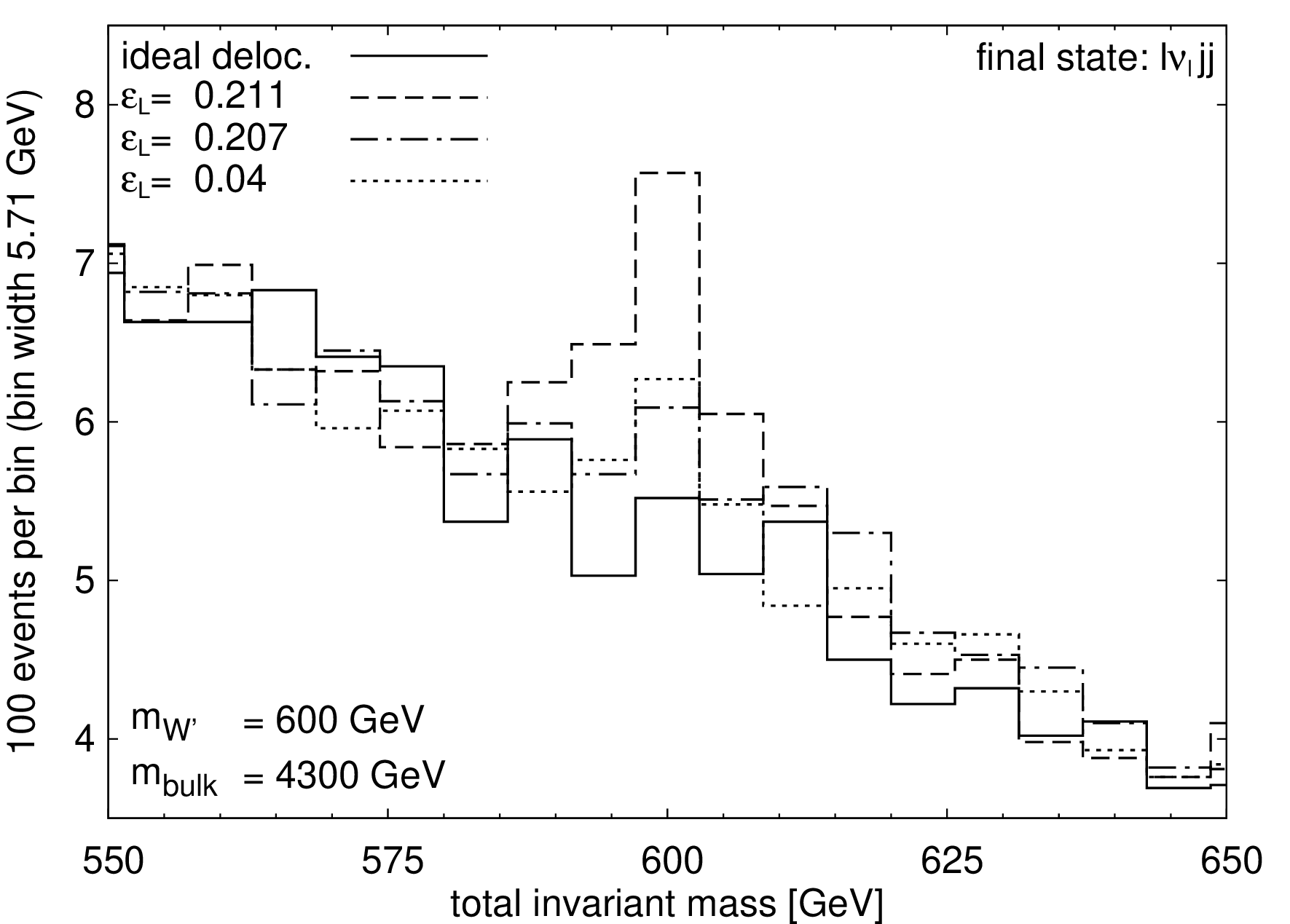} &
\includegraphics[angle=270,width=6.5cm]{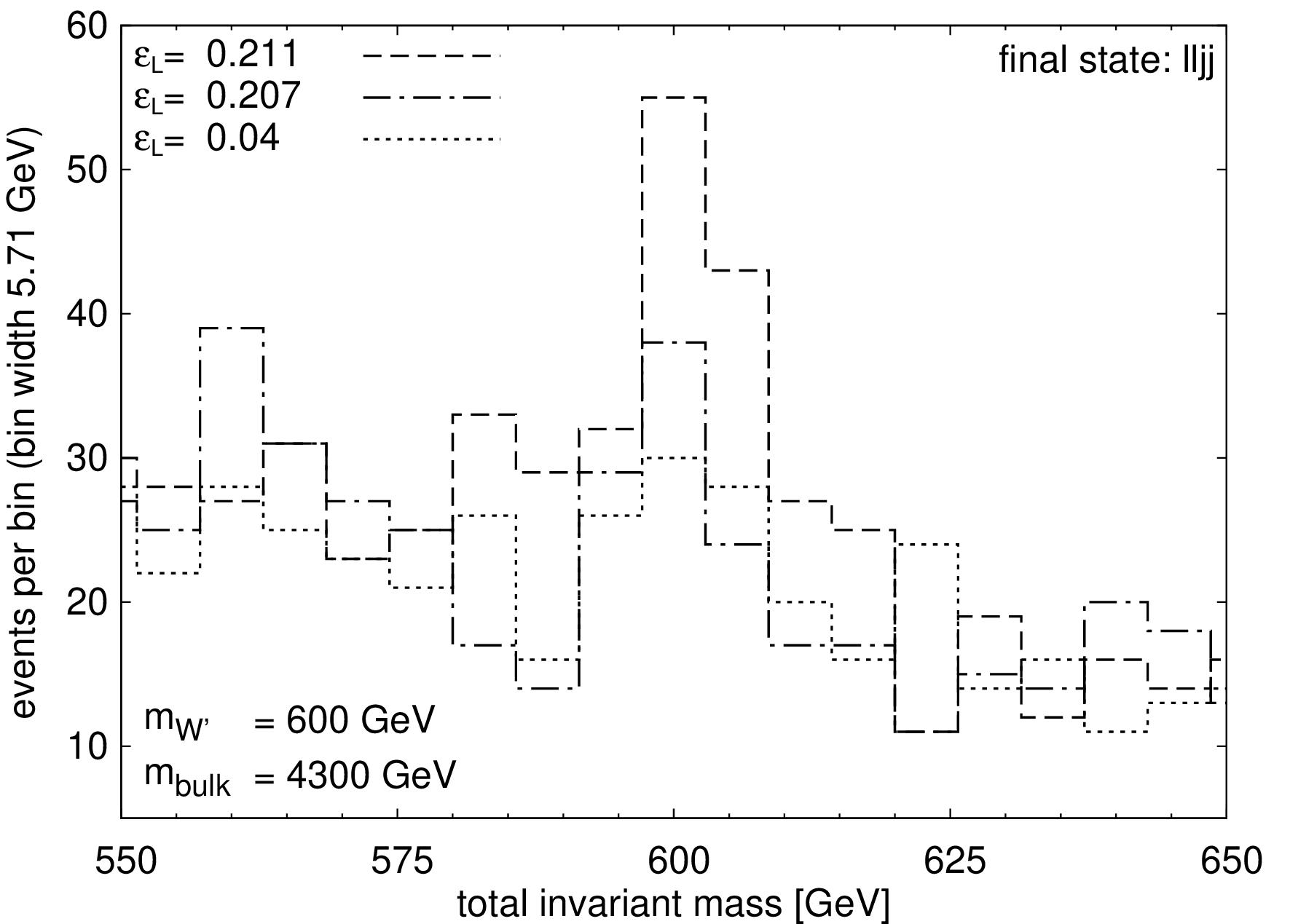}
\end{tabular}}
\caption{\emph{Left column: }The $W^\prime$ resonance peak in the invariant mass
distribution for $pp\rightarrow l\nu_ljj$ for different values of the delocalization
parameter. \emph{Right column: }The same distributions in the case of the $lljj$ final
state.}
\label{hist-hw-deloc}
\end{figure}
The dependence of the resonance peak on the delocalization parameter $\epsilon_L$ is shown
in figure~\ref{hist-hw-deloc}. The left column shows the $\pm\unit[50]{GeV}$ region around
the peak for the $l\nu_ljj$ final state for different values of $\epsilon_L$ 
and for the case of ideal delocalization. For
$m_{W^\prime}=\unit[500]{GeV}$ and $m_{W^\prime}=\unit[600]{GeV}$ the peak vanishes in the
case of ideal delocalization which demonstrates that the cut~(\ref{equ-cut-wzident}) is
sufficient to discriminate between jets coming from the decay of $W$ and those coming from
a $Z$. In the case of
$m_{W^\prime}=\unit[380]{GeV}$, a small peak remains even in the case of ideal
delocalization which stems from jets coming from
$pp\rightarrow Z^\prime\rightarrow l\nu_ljj$ misidentified as a $Z$ (we will discuss the
possibility of unfolding these two contribution in the next section).

The histograms show
that tuning $\epsilon_L$ towards the point of ideal delocalization quickly decreases the
size of the peak making it invisible for the lowest chosen values of~$\epsilon_L$.
The right column shows the same region around the peak for the final state
$lljj$ and the same values of $\epsilon_L$. As should be expected, the same decrease of
the peak size is visible.

To obtain a numerical estimate for the integrated luminosity required
for a $s=5\sigma$ or~$3\sigma$
discovery of the $W^\prime$ at some given value of the delocalization
parameter $\epsilon_L$ we exploit the fact that the significance of
the signal scales as~$g_{W'ff}^2$
with the coupling of $W^\prime$ to left-handed SM fermions. This allows
us to estimate the integrated luminosity required for
obtaining a signal with significance $s_0$ in terms of the significance of the signal
for other values of coupling and integrated luminosity.

For the actual determination of $s$ from Monte Carlo data we define the signal as in section
\ref{sec-hzprod}. In case of the $l\nu_ljj$ final state we calculate~$s$
via~(\ref{equ-sgn-rec}), while for the case of $lljj$ it can be calculated simply as
\begin{equation}
  s = \frac{N_s}{\sqrt{N_b}}\,,
\end{equation}
because we don't have the additional doubling of the
background events by the neutrino momentum reconstruction in this case\footnote
{Because of the lower number of events in the final state $lljj$, the background for this
case was calculated for an integrated luminosity of $\int\mathcal{L}=\unit[1000]{fb^{-1}}$
and scaled down.}.

\begin{table}
\centerline{
\begin{tabular}{|c|c|c||c|}
\hline\multicolumn{4}{|c|}{$W^\prime\rightarrow l\nu_ljj$} \\\hline\hline
$m_{W^\prime}\:\left[\unit{GeV}\right]$ & $m_\text{bulk}\:\left[\unit{TeV}\right]$
& $\epsilon_L$ & $s$ \\
\hline 380 & 3.5 & 0.338 & 5.6 \\
\hline 500 & 3.5 & 0.254 & 8.6 \\
\hline 600 & 4.3 & 0.211 & 6.4 \\ \hline
\end{tabular}\hspace{1cm}
\begin{tabular}{|c|c|c||c|}
\hline\multicolumn{4}{|c|}{$W^\prime\rightarrow lljj$} \\\hline\hline
$m_{W^\prime}\:\left[\unit{GeV}\right]$ & $m_\text{bulk}\:\left[\unit{TeV}\right]$
& $\epsilon_L$ & $s$ \\
\hline 380 & 3.5 & 0.338 & 5.1 \\
\hline 500 & 3.5 & 0.254 & 8.5 \\
\hline 600 & 4.3 & 0.211 & 8.3 \\ \hline
\end{tabular}}
\caption{The significance of the signal calculated at different points in parameter space
for both final states.}
\label{tab-sgn-hw}
\end{table}
The significances calculated this way at different points in parameter
space are shown in
table~\ref{tab-sgn-hw}. For $m_{W^\prime}=\unit[380]{GeV}$ and
$m_{W^\prime}=\unit[500]{GeV}$, both final states seem to do equally well at revealing the
fermionic couplings of the $W^\prime$; however, for $m_{W^\prime}=\unit[600]{GeV}$ the
dilepton final state appears to give a slightly better signal owing to the
better ratio of signal to background.
\begin{figure}
\centerline{\begin{tabular}{cc}
\multicolumn{2}{c}{\includegraphics[angle=270,width=6.5cm]{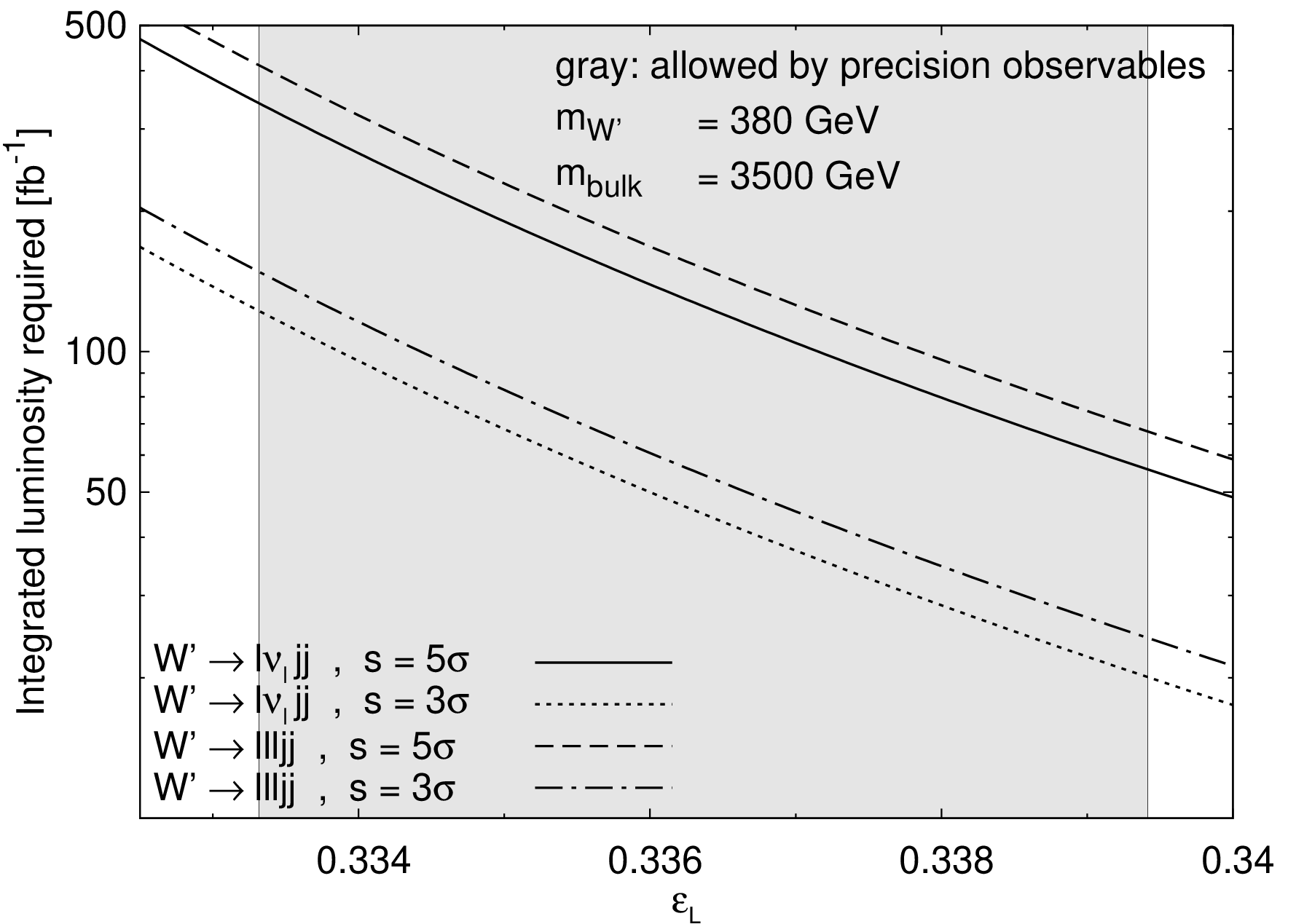}} \\
\includegraphics[angle=270,width=6.5cm]{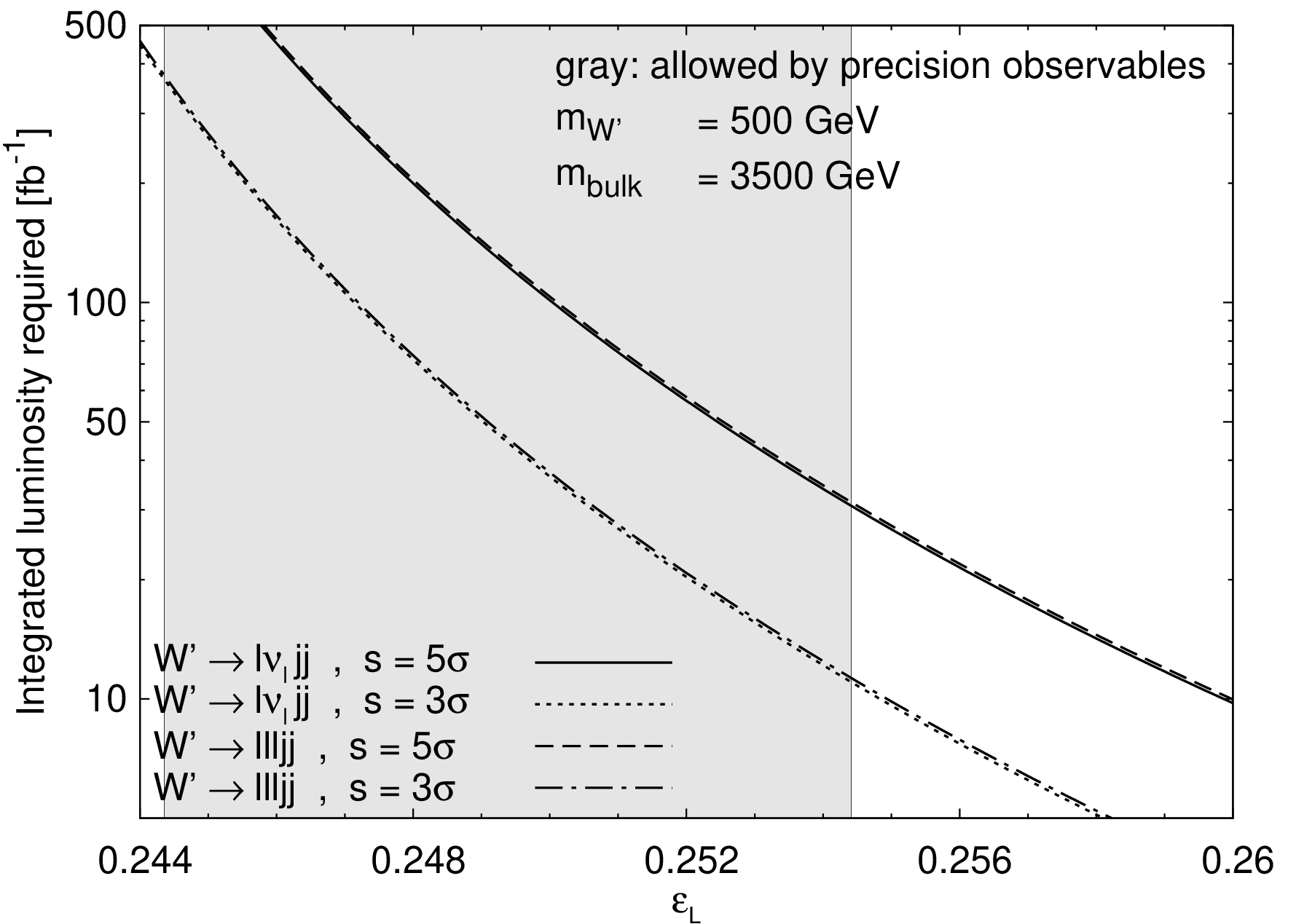} &
\includegraphics[angle=270,width=6.5cm]{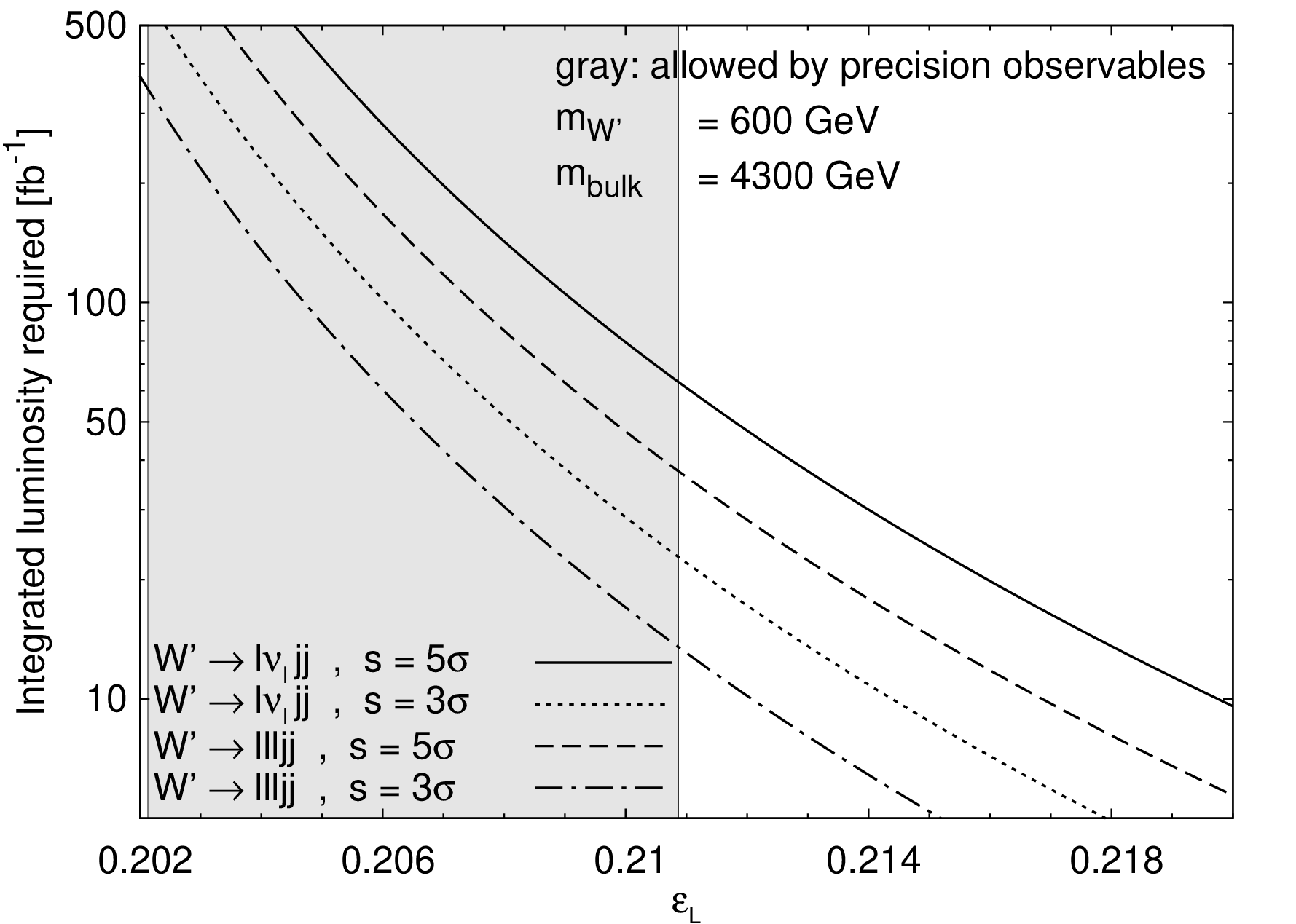}
\end{tabular}}
\caption{The integrated luminosity required for a $5\sigma$ resp. $3\sigma$ discovery of the
$W^\prime$ in the $s$-channel.}
\label{fig-lumi-hw}
\end{figure}
The integrated luminosity necessary for a $5\sigma$ resp. $3\sigma$ discovery of the
$W^\prime$ in the $s$-channel is shown in figure~\ref{fig-lumi-hw} together with the range of the
delocalization parameter $\epsilon_L$ allowed for the different choices of $m_{W^\prime}$
and $m_\text{bulk}$. Taking the integrated luminosity collected over the full LHC running
time to be around $\unit[400]{fb^{-1}}$ and considering the fact that the band of allowed
$\epsilon_L$ (and $g_{W^\prime ff}$) can be moved further towards smaller values by
lowering $m_\text{bulk}$, it is evident from figure~\ref{fig-lumi-hw} that there is a part of
the allowed parameter space in which the $W^\prime$ would appear perfectly fermiophobic
at the LHC. However, there also is a big region of parameter space in which the coupling of
the $W^\prime$ to the SM fermions eventually should be discovered, although
this still would take several years of running time as the lowest integrated luminosity
required for $3\sigma$ is around $\unit[10]{fb^{-1}}$ even at the point in parameter space
most easily accessible.

\section{Finite jet resolution and $W$/$Z$ identification}
\label{sec-jetres}

Since flavor tagging is impossible for light quark flavors, we have to
rely on invariant mass cuts for the jet pairs to be able to separate
the case of the two jets in $l\nu_ljj$ coming from the decay of a~$W$
in $Z^\prime$~production from that of the jets being produced by a
decaying $Z$ in $W^\prime$~production.
However, it may very well be impossible to obtain a resolution of order
$\pm\unit[5]{GeV}$ in the jet invariant mass from experimental data.
In this section, we discuss the effect of a gaussian smearing of
the invariant mass of the jets on our analysis.

In the ideal case of exact $m_{jj}$ measurement, events coming from the decay of a
intermediary $W$/$Z$ are distributed according to a Breit-Wigner distribution
\[ p_b(x,m,\Gamma)\:dx =
\frac{n_b(m,\Gamma)^{-1}}{\left(x^2-m^2\right)^2+\Gamma^2 m^2}\:dx\,, \]
with the normalization factor
\[ n_b(m,\Gamma) = \frac{\pi}{4m^3}\left(1+\frac{\Gamma^2}{m^2}\right)^{-\frac{3}{4}}
\sin^{-1}\left(\frac{1}{2}\atan\frac{\Gamma}{m}\right)\,. \]
Emulating the measurement error in the jet mass by convoluting $p_\text{bw}$ with a
gaussian of standard deviation $\sigma$
\[ 
p_\text{g}(x,\sigma)\:dx = \frac{1}{\sqrt{2\pi}\sigma}e^{-\frac{x^2}{2\sigma^2}}\:dx
\]
we obtain the smeared distribution
\[ p_\text{sm}(x,m,\Gamma,\sigma)\: dx = \int_0^\infty dy\:p_b(y,m,\Gamma)
p_\text{g}(x-y,\sigma)\,.
\]
\begin{figure}
\centerline{\includegraphics[angle=270,width=8cm]{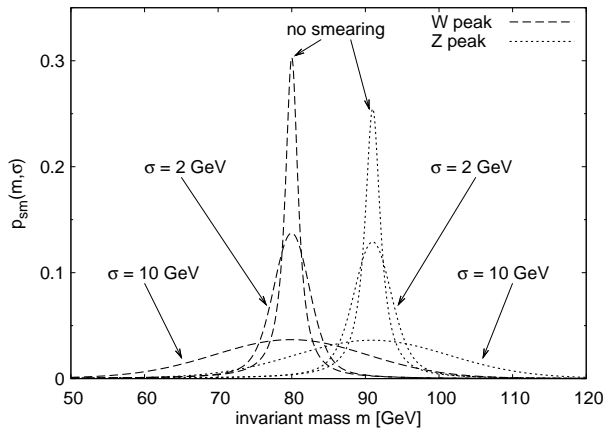}}
\caption{The effect of a gaussian smearing on the Breit-Wigner shape of the $W$ and $Z$
resonances for various widths $\sigma$ of the gaussian.}
\label{fig-smearedpeaks}
\end{figure}

Figure~\ref{fig-smearedpeaks} shows the effect of this smearing on the Breit-Wigner peaks of
the $Z$ and the $W$. Turning on the smearing and increasing $\sigma$ causes the sharp
Breit-Wigner peaks to decay rapidly, and for $\sigma=\unit[10]{GeV}$, only two very broad
bumps are left. The result is that, if a cross section has one contribution which stems
from the decays of a virtual $Z$ and one coming from a virtual $W$, any attempt to isolate
the $Z$ contribution by cutting on the resonance will inevitably also select events coming
from the $W$ decay contaminating the sample (and vice versa). Therefore, our analysis of
the $l\nu_ljj$ final state will show a $W^\prime$ peak even in the case of ideal
delocalization which is caused by jet pairs from a decaying $W$ misidentified as a $Z$.

If we try to isolate the $W$ peak with a cut on the invariant mass $m_{jj}$
\[ L_W \le m_{jj} \le U_W \]
and the $Z$ peak with a cut
\[ L_Z \le m_{jj} \le U_Z\,, \]
then the resulting event counts $\widetilde{N}_W, \widetilde{N}_Z$ can be
calculated from the true event counts $N_W, N_Z$  coming from a decaying $W$ or $Z$
via a matrix $T$ as
\[ \begin{pmatrix} \widetilde{N}_W \\ \widetilde{N}_Z \end{pmatrix} =
\begin{pmatrix} T_{WW} & T_{WZ} \\ T_{ZW} & T_{ZZ} \end{pmatrix}
\begin{pmatrix} N_W \\ N_Z \end{pmatrix} \]
with entries
\[ T_{ij} = \int_{L_i}^{U_i}dm\:p_\text{sm}(m,m_j,\Gamma_j,\sigma)\,. \]
Inverting $T$ we can calculate the event counts $N_W$ and $N_Z$
\begin{equation} \begin{pmatrix} N_W \\ N_Z \end{pmatrix} = T^{-1}
\begin{pmatrix} \widetilde{N}_W \\ \widetilde{N}_Z \end{pmatrix}\,.
\label{equ-trans-mat}\end{equation}
The entries of $T$ give the probability of misidentifying an event and can be readily
calculated numerically; for example, choosing cuts
\[ L_W=\unit[60]{GeV} \quad,\quad U_W=\unit[85]{GeV} \quad,\quad
L_Z=\unit[86]{GeV} \quad,\quad U_Z=\unit[111]{GeV} \]
yields
\[ T \approx \begin{pmatrix} 0.64 & 0.27 \\ 0.29 & 0.62 \end{pmatrix}
\quad,\quad
T^{-1} \approx \begin{pmatrix} 1.9 & -0.85 \\ -0.89 & 2.0 \end{pmatrix}\,. \]
This way, we can in principle use $T$ to disentangle the
contributions from $W$ and $Z$ resonances
to the signal in the presence of a measurement error which causes the Breit-Wigner
peaks to lose their shape. However, to apply this to actual data, it is vital
to separate the signal from both the reducible and the irreducible
backgrounds, because they don't follow a Breit-Wigner distribution.

In order to estimate the significance of a signal obtained this way, we calculate the
standard deviation $\sigma_{N_i}$ of $N_i$ according to
\[ \sigma_{N_i} = \sqrt{\sum_{j\in W,Z}\left(T^{-1}_{ij}\right)^2\sigma_{\widetilde{N}_j}^2}\,. \]
In our analysis, we obtain the signal events inside the smeared Breit-Wigner peaks
$\widetilde{N}_i$ by subtracting the background $N_{b,i}$ from the total number of events
$N_{t,i}$. The error on $N_{t,i}$ is
\[ \sigma_{N_{t,i}} = \sqrt{N_i + 2N_{b,i}} = \sqrt{N_{t,i} + N_{b,i}}\,, \]
because of the neutrino momentum reconstruction doubling the amount of background events
(cf.~section~\ref{sec-hzprod}), and we finally arrive at
\begin{equation}
\sigma_{N_i} = \sqrt{\sum_{j\in W,Z}\left(T^{-1}_{ij}\right)^2\left(N_{t,j} + N_{b,j}\right)}
\label{equ-sigma-after-transfer}\end{equation}
For a simulation of the effect of the measurement error our analysis we have
randomly distributed the invariant mass of the jet pairs within a gaussian with width
$\sigma=\unit[10]{GeV}$ centered around the correct value calculated from Monte Carlo data. We
then did the same analysis as in sections \ref{sec-hzprod} and \ref{sec-hwprod}
with $m_{W^\prime}=\unit[500]{GeV}$ and $m_\text{bulk}=\unit[3.5]{TeV}$ both for
$\epsilon_L=0.254$ and for the ideally delocalized scenario. The only difference to the
previous analysis are the cuts on $m_{jj}$ which we enlarged to
\[ \unit[60]{GeV}\le m_{jj}\le\unit[85]{GeV} \quad\text{resp.}\quad
\unit[86]{GeV}\le m_{jj} \le\unit[111]{GeV}\,. \]
\begin{figure}
\centerline{\begin{tabular}{cc}
\includegraphics[angle=270,width=6.5cm]{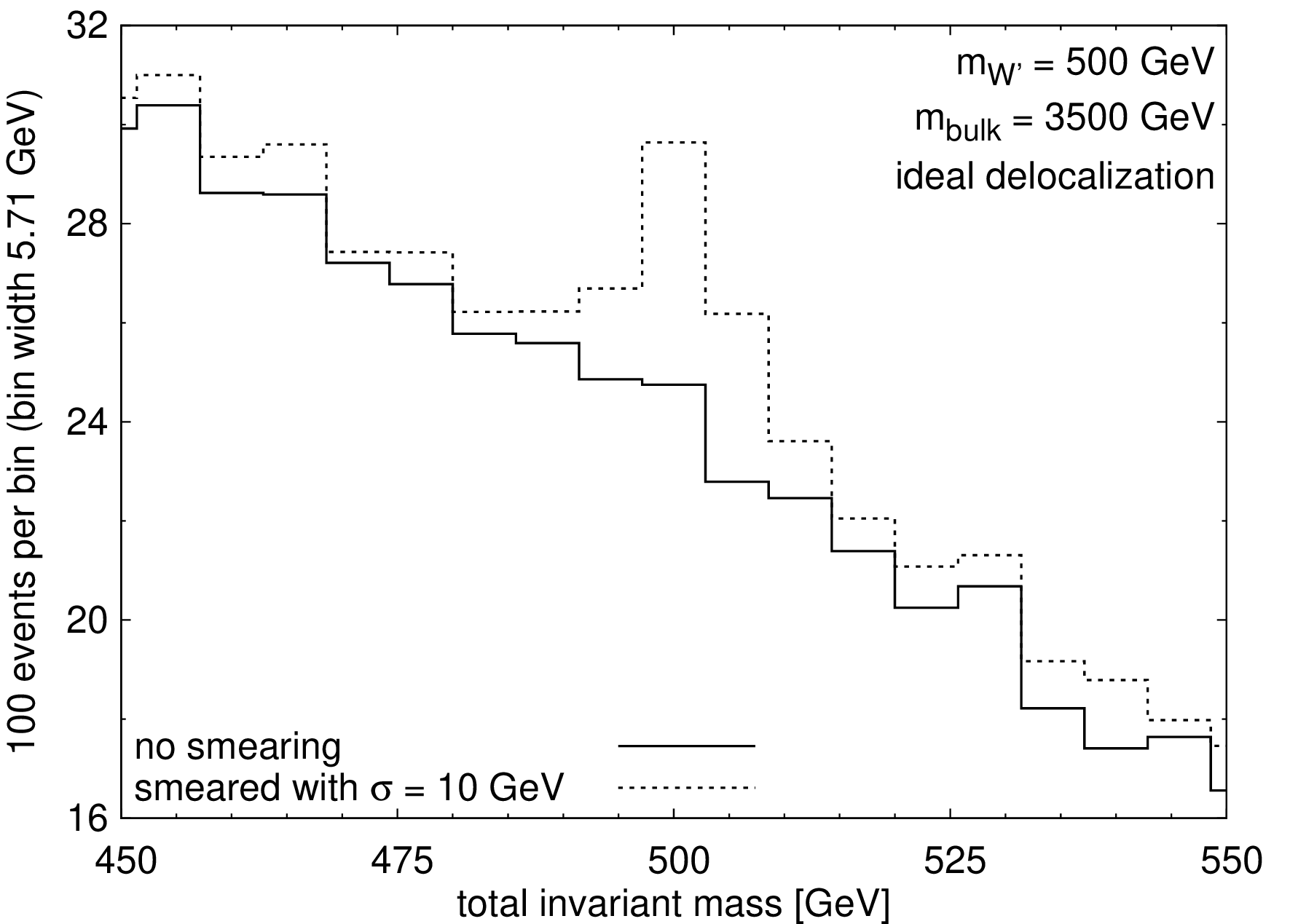} &
\includegraphics[angle=270,width=6.5cm]{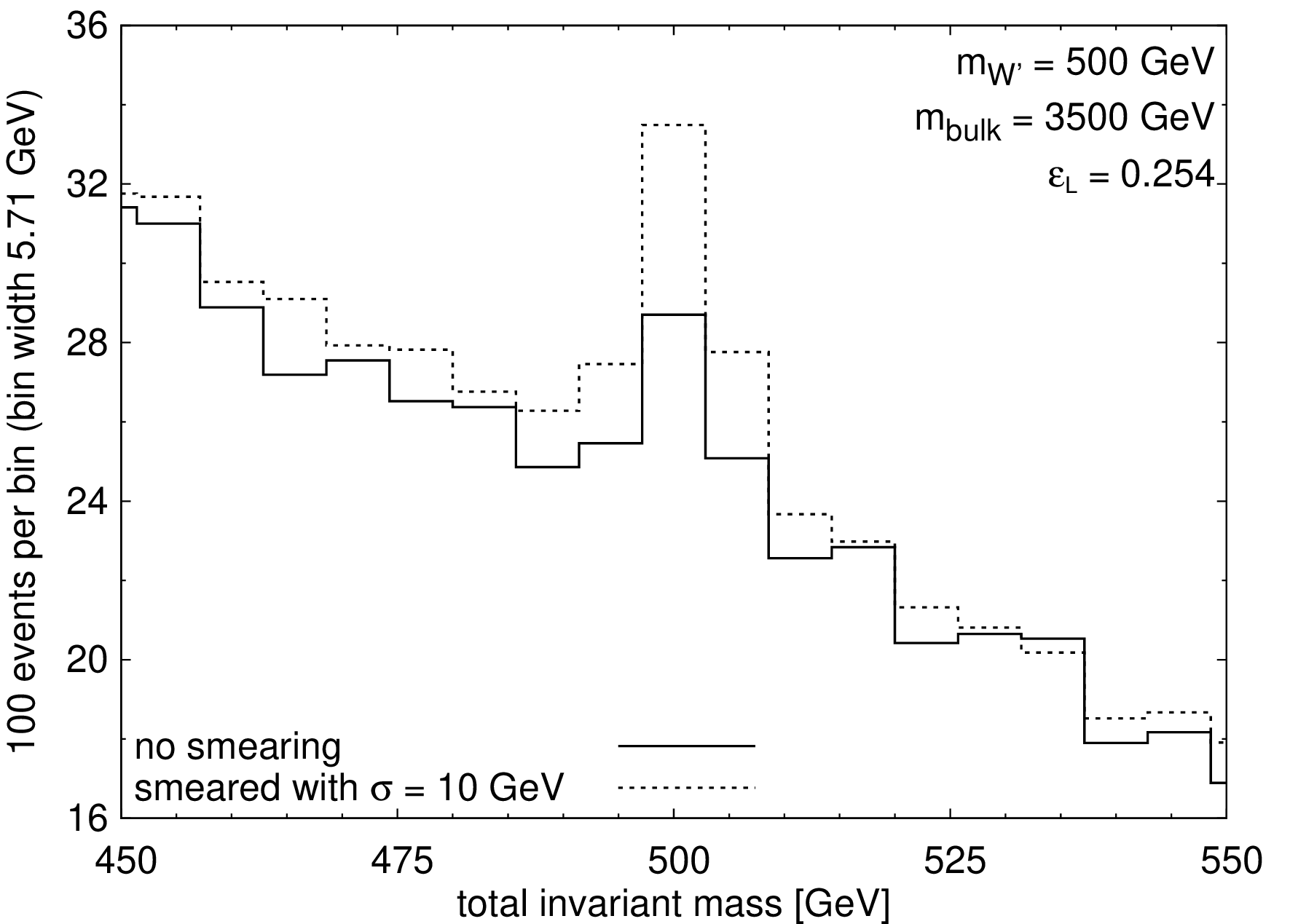}
\end{tabular}}
\caption{\emph{Left: }Signal in the $W^\prime$ detection channel for the case of ideal
delocalization smeared with a gaussian error. \emph{Right: }The same for the case
of nonzero $g_{W'ff}$}
\label{hist-wsmear}
\end{figure}
Figure~\ref{hist-wsmear} shows the resulting effect on the $W^\prime$ peak for the cases
of ideal delocalization (left) and for $\epsilon_L=0.254$ (right). In both cases a peak is
clearly visible, which in the ideally delocalized scenario is only composed of events
with jets coming from a decaying $W$ misidentified as a $Z$.

The number of signal events
$\widetilde{N}_{W/Z}$ after smearing, the significance $s_{W/Z}$ of these
calculated via~(\ref{equ-sgn-rec}), $N_{W/Z}$ obtained from applying the transfer
matrix $T^{-1}$ (\ref{equ-trans-mat}) and the resulting significance
$N_i/\sigma_{N_i}$ obtained from~(\ref{equ-sigma-after-transfer}) are shown in
table \ref{tab-sgn-wzsep}. All peaks are significant with $s>5\sigma$; however, after
applying the transfer matrix, the $W^\prime$ peak vanishes within one standard deviation
for ideal delocalization,
while in the case of $\epsilon_L=0.254$ a residue as big as $2\sigma$ remains. The
$Z^\prime$ peak remains significant after applying the transfer matrix, however, the
significance is reduced because the transfer matrix enlarges the error.
\begin{table}
\centerline{
\begin{tabular}{|c||c|c|c|c|}
\hline\multicolumn{5}{|c|}{ideal delocalization}\\\hline\hline
 & $\widetilde{N}_i$ & $s_i$ & $N_i$ & $\frac{N_i}{\sigma_{N_i}}$ \\\hline\hline
$i=W$ & $3193$ & $17$ & $5126$ & $13$ \\\hline
$i=Z$ & $1371$ & $7.5$ & $-96.10$ & $0.24$ \\\hline
\end{tabular}
\hspace{1cm}
\begin{tabular}{|c||c|c|c|c|}
\hline\multicolumn{5}{|c|}{$\epsilon_L=0.254$}\\\hline\hline
 & $\widetilde{N}_i$ & $s_i$ & $N_i$ & $\frac{N_i}{\sigma_{N_i}}$ \\\hline\hline
$i=W$ & $3767$ & $21$ & $5628$ & $14$ \\\hline
$i=Z$ & $2083$ & $11$ & $811.6$ & $2.0$ \\\hline
\end{tabular}
}
\caption{Comparison of the signals $\widetilde{N}_{W/Z}$ obtained with an gaussian
smearing of the invariant mass of the jets with $\sigma=\unit[10]{GeV}$ to the ``true''
signals $N_{W/Z}$ calculated from the measured ones via the transfer matrix $T^{-1}$.}
\label{tab-sgn-wzsep}
\end{table}

What are the consequences for the detection of $Z^\prime$ and $W^\prime$ in the $l\nu_ljj$
final state? The detection of the $Z^\prime$ is not affected by inaccuracies in the
jet mass resolution as the peak is always present with little variations of its size over
the whole parameter space, and we can always compensate for the smearing of the jet
mass by enlarging the cut window on $m_{jj}$. However, the separation of a possible $W^\prime$
contribution to the peak (which depends heavily on the point in parameter space) by
cutting on $m_{jj}$ alone is spoiled by the error in $m_{jj}$; we have to apply additional
tricks like the transfer matrix~(\ref{equ-trans-mat}) to disentangle the two contributions.
While this seems to work in principle, the significance of the $W^\prime$ signal is
reduced by this analysis, rendering this final state much less suitable for detecting a
coupling between $W^\prime$ and SM fermions than the decay into $lljj$ which
is not contaminated by a contribution of the $Z^\prime$.

\section{Conclusions}
\label{sec:concl}

We have studied the production of the heavy~$W'$ and $Z'$ bosons of
the three site higgsless model in the $s$-channel at the LHC.  Unlike
vector boson fusion, this production mode allows to directly measure
the couplings of the new bosons to standard model fermions.  These
couplings are constrained by electroweak precision tests and their
measurement is therefore crucial for consistency checks of models of
electroweak symmetry breaking with extended gauge sectors.

We have found a method that will allow the separation of~$W'$ from
$Z'$~processes at the parton level.  Our results show that the
observation of $s$-channel production of $Z'$ bosons will not require
a lot of integrated luminosity for all of the allowed parameter space.
In contrast, $W'$ production in the $s$-channel is much more sensitive
to the model parameters and there are regions of parameter space where
an observation will be very challenging, if not impossible.  A more
detailed experimental analysis should investigate the effects of
hadronization and detector response on our results.

\section*{Acknowledgments}
This research is supported by Deutsche Forschungsgemeinschaft through
the Research Training Group 1147 \textit{Theoretical Astrophysics and
Particle Physics}, by Bundesministerium f\"ur Bildung und Forschung
Germany, grant 05HT6WWA and by the Helmholtz Alliance \textit{Physics
at the Terascale}.


\end{document}